\documentclass{article}

\usepackage{PRIMEarxiv}
\usepackage[utf8]{inputenc} 
\usepackage[T1]{fontenc}    
\usepackage{hyperref}       
\usepackage{url}            
\usepackage{booktabs}       
\usepackage{amsfonts}       
\usepackage{nicefrac}       
\usepackage{microtype}      
\usepackage{lipsum}
\usepackage{fancyhdr}       
\usepackage{graphicx}       
\graphicspath{{media/}}     

\usepackage{soul}
\usepackage{graphicx,subcaption}
\usepackage{algorithm} 
\usepackage{algorithmic}
\title{PhysiQ: Off-site Quality Assessment of Exercise in Physical Therapy
\thanks{\href{https://doi.org/10.1145/3570349}{https://doi.org/10.1145/3570349}}
}

\author{
  Hanchen David Wang \\
  Vanderbilt University \\
  hanchen.wang.1@vanderbilt.edu \\
  \And
  Meiyi Ma \\
  Vanderbilt University \\
  meiyi.ma@vanderbilt.edu
  }

\begin{document}
\maketitle


\begin{abstract}
Physical therapy (PT) is crucial for patients to restore and maintain mobility, function, and well-being. Many on-site activities and body exercises are performed under the supervision of therapists or clinicians. However, the postures of some exercises at home cannot be performed accurately due to the lack of supervision, quality assessment, and self-correction. Therefore, in this paper, we design a new framework, PhysiQ, that continuously tracks and quantitatively measures people's off-site exercise activity through passive sensory detection. In the framework, we create a novel multi-task spatio-temporal Siamese Neural Network that measures the absolute quality through classification and relative quality based on an individual's PT progress through similarity comparison. PhysiQ digitizes and evaluates exercises in three different metrics: range of motions, stability, and repetition. 
We collect and annotate 31 participants’ motion data with different levels of quality. 
Evaluation results show that PhysiQ recognizes the nuances in exercises, works with different numbers of repetitions, and achieves an accuracy of 89.67\% in detecting levels of exercise quality and an average R-squared correlation of 0.949 in similarity comparison. 
\end{abstract}


\maketitle
\section{Introduction}
\label{sec:introduction}
{Physical therapy} (PT) is a process in which patients regain their strength through exercises after surgery, incidents, or illness. It benefits patients by reducing pain, improving mobility, preventing further injury, and improving muscle balance. Patients undergo challenges, endeavors, and struggles with lasting benefits with well-prescribed instruction and supervision. 
Currently, there are more than 5.1 billion Years Lived with Disability (YLDs)\footnote{This measures the impact of an illness before it resolves or leads to death} growth per year \cite{jesus2019global}.  In the U.S., there are 38,800 physical therapy clinics operating and an average of 150 patients in each clinic each week, with approximately 300 million sessions for patients each year \cite{salazar_2019}. Usually, patients require extensive sets of exercises to return to regular activities. However, they usually have limited time in clinics with supervision under physical therapists, and are required to perform exercises by themselves \textbf{at home}. 
Nevertheless, patients have very little knowledge of how well they perform without monitoring or supervision. Moreover, they do not have the flexibility and convenience to set up a camera to self-monitor~\cite{stankovic2021challenges}. Therefore, having a quantitative measurement of the quality of exercises with wearable devices for patients and therapists is crucial to support the patients get their wellness back.

\subsection{Motivation}
To improve effective rehabilitation, self-efficacy, self-motivation, social support, intentions, and previous adherence to physical therapies can help patients perform exercises in home-based physical therapy \cite{essery2017predictors}. 
However, there is a considerable gap existing between how patients perform self-monitored offsite therapeutic and clinically supervised exercises. Patients and their therapists have no effective and convenient way to track exercises quantitatively at home.

Human Activity Recognition (HAR) in wearable devices is a prevalent research topic that includes many day-to-day locational, behavioral, and planning recognition. For instance, handwashing \cite{wang2021you}, finger gestures \cite{chen2021vifin}, eating behavior \cite{bi2018auracle}, and daily activities (writing, cooking, and cleaning) \cite{bhattacharya2022leveraging} improve to recognize activities through wearable technologies. However, although these works meet the need for daily human activities, a limited attempt exists to help therapeutic rehabilitation for patients. 

Moreover, existing works on the quality of exercise utilize vision-based devices, such as cameras and K2 Kinect \cite{NESHOV2019, HAGHIGHIOSGOUEI2020}, to track the quality of exercises and provide simple feedback. However, users must set them up physically and potentially interfere with the occlusion of the camera. Moreover, calibrating and adjusting vision-based devices are costly and inconvenient for injured or immobile people. Therefore, there is a need to provide portable and wearable devices to measure the quality of exercises.

Lastly, attempts on wearable devices to track quality, such as calorie intake \cite{hussain2022smart} and gait authentication \cite{papavasileiou2021gaitcode}, suggest the endeavors to use wearable sensors to track the quality of exercises. Ghanashyama et al. attempt to use deep learning models to recognize and count the repetitions of exercises using a single sensor \cite{PRABHU2021}. However, qualitative information is not a therapeutic metric to help rehabilitation. 

In summary, there is a high demand to improve how patients and users quantitatively measure their exercises offsite with meaningful feedback from wearable devices.

\subsection{Challenges}
There are three significant challenges in designing such a framework. 
First, how to digitize physical metrics is an open question.
To the best of our knowledge, there are no existing systems or models measuring the quality of an activity, and existing models are not sophisticated enough to directly return a quantitative measurement.  
Secondly, the quality of exercise varies for different people of different ages, heights, weights, and gender. For example, a tall person has a long traveling distance for shoulder abduction exercise because of his height and arm lengths. Suppose someone of average height performs the same PT exercise with the same range of motion, the model is supposed to be able to tell the difference and similarities even though their heights are different. Similarly, suppose an elder performs differently than a teenager in a PT exercise; even though their objective quality differs, quality should remain the same if they raise their arm to 90 degrees compared to 60 degrees.
Lastly, there are no existing datasets with annotated quality of exercises.

\subsection{Contributions}
Targeting these challenges, this paper introduces a novel framework, PhysiQ, that continuously tracks and quantitatively measures people's off-site exercise activity through passive sensory detection. 
The framework is robust and general to handle users who perform exercises with different speeds, positions, and postures. We summarize our main contributions below:
\begin{itemize}
    \item To the best of our knowledge, this is the first framework for quantitative measurement of exercises using a smartwatch.  
    The framework identifies and digitalizes three key exercise metrics of \textit{range of motion}, \textit{stability}, \textit{repetition}, which are built upon the muscular system for understanding the functionality of the skeletal system.
    \item We create a novel multi-task spatio-temporal Siamese Neural Network that  
    measures both absolute quality and relative quality based on an individual's PT progress through similarity comparison. It enables patients to understand the quality of their offsite exercises over time.
    \item We build an application collecting users' motion data in a smartwatch and giving explainable feedback with recommendations based on their quality of exercises in real-time. 
    \item We collect and annotate 31 participants' motion data with different levels of stability, range of motion, and repetition, in three shoulder exercises, which are shoulder abduction, external rotation, and forward flexion.
    \item We perform an extensive evaluation using real user's data. Results show that our framework performance outperforms the baselines by 47.67\% on average in R-Squared for all exercises and all three metrics.  
    We also provide insights of how user's behaviors influence the framework through a user experience study. 

\end{itemize}

\subsection{Paper Organization}
In the rest of the paper, we discuss the related work in  Section \ref{sec:related work}. We present our framework PhysiQ in  Section \ref{sec:solutions}. Next, we show how we collect the exercise data with different quality in  Section \ref{sec:data collection}, and evaluation results in  Section \ref{sec:evaluation}. Furthermore, we present a survey and discuss user experience using our app and implications in  Section \ref{sec:survey}, followed by a discussion and summary in  Section \ref{sec:discussion} and  Section \ref{sec:summary}, respectively.

\section{Related Work}
\label{sec:related work}
In this section, we present existing literature on state-of-the-art measuring the quality of activity and applications, and deep learning methods for these applications through similarity comparison and other methods.

\subsection{Measuring Quality of Activity}
Though there does not exist a field of such, Human Activity Quality Recognition (HAQR) is a critical research question for real-world application. Therefore, we have gathered and scrutinized related works in state-of-the-arts. For example, one quality in exercises is repetition counting. Work, such as \cite{stromback2020mm}, focuses on multi-modality to provide a more accurate result of repetitions counting and exercise recognition. A fascinating work done by Radhakrishnan et al. focuses on using in-ear devices such as wireless headphones fusing with inertial measurement units to quantify insights and feedback in gym exercises \cite{radhakrishnan2019can}.

Additionally, one work uses smart speakers to analyze the duration, intensity, continuity, and smoothness of exercises at home \cite{xie2021hearfit}. However, such work requires people to have knowledge about exercises and some level of proximity to the actual devices. Additionally, IMUTube introduces how to simulate virtual IMU data from video \cite{kwon2020imutube}. Kwon et al. utilizes video to simulate IMU through a number of off-the-shelf computer vision and graphics techniques. Furthermore, Radhakrishnan et al. suggest a system that uses a magnetic accelerometer sensor. The device is mounted on the weight stack of a gym machine to infer exercise behavior using multiple machine learning models to identify the person, amount of weights, type of exercise, and mistakes \cite{radhakrishnan2021w8}.

Like the quality of exercises, sleep quality analyzes humans' brain activity through Electroencephalography (EEG) signal. It categorizes the level of sleep activity in Rapid Eye Movement (REM), Non-REM, S2 (light sleep), and S3 (slow-wave sleep). Additionally, polysomnography is considered the standard methodology for detecting the sleep pattern with carefully analyzed records of epochs \cite{crivello2019meaning}. There are two leading technologies for sleep monitoring: 
portable and contact. Portable devices such as smartwatches or smartphones seamlessly collect users' data through passive means. Contact devices are medical-level tools to collect reliable data through less passive means \cite{crivello2019meaning}. 
 
Portable devices require stationary devices to record the daily routine of the subjects \cite{sleep:chang2018sleepguard, sleep:mehrabadi2020sleep, sleep:scherz2017heart, sleep:sun2017sleepmonitor,ma2017m}. These works include signals of accelerometer data to approximate respiration rate and heart rate through, for example, sound recording to validate sleep time and duration, and electrocardiogram (ECG) signal to proximate heart rate and distinguish threshold for sleeping and waking. Contact devices, which require direct contact with the subjects, are the cases of PPG, EEG, and Actigraphy. These assessment technologies have the main advantage of accurately sampling the human body's physiological and mechanical signals. However, such devices are perceived as obtrusive due to their limited portability and transparency. Sleep quality recognition and detection using contact devices have been well studied. One research targets whether having additional information such as age and sleep stage information can distinguish abnormality using the deep learning method on EEG signal data \cite{sleep:contact:van2019detecting}. Other research focuses on developing an automatic sleep staging method in EGG signals, in which the author proposes multi-epoch methods to segment and encodes the feature and re-concatenate to predict its sleep stage \cite{sleep:contact:li2021end}. Lastly, one aims to develop a sleep scoring toolbox with the competency of multi-signal processes, feature extraction, and classification and prediction but only using a simple logic-feature-based method to differentiate sleep stage \cite{sleep:contact:yan2019automatic}.


\subsection{Deep Learning Models for Quality Measurements} 
Siamese Neural Network (SNN), utilizing two identical networks, is commonly applied to compare if two subjects are same or not. 
Typical tasks include image classification, object recognition, and object tracking \cite{he2018twofold, dong2018triplet, shen2019visual, wang2018learning, guo2017learning, leal2016learning}. However, it only returns a binary result (i.e., True or False) without a quantitative measurement.  

Furthermore, recognizing similarities in 1-D signal data, such as radar, speech, and natural language process (NLP) \cite{govalkar2021siamese, mittag2020full, neculoiu2016learning, benajiba2019siamese}, has also gained many usages in different applications. 
An intriguing application stands out using semantic similarity between sentences, supplementing recurrent neural networks with synonymic encoding. Mueller et al. use LSTM to encode the different length inputs with its positional encoder to analyze the semantic similarity with an outstandingly high Pearson correlation of 0.8822 \cite{mueller2016siamese}. Furthermore, A deep dive evaluation of the SNN proposes a residual module to reduce learning biases caused by padding \cite{zhang2019deeper}. The determinant factors, such as stride, padding, and receptive field size (the size of the region that produces the feature), are crucial to the performance of the SNN. Additionally, Lawrence et al. suggest an innovative way of utilizing spatial and temporal aspects from videos to recognize human emotions is inspiring. \cite{lawrance2021emotion}. Lastly, Agrawal et al. use an inventive way to calculate articles' similarity distance for political stance using discrete labeling of relatedness \cite{agrawal2017cosine}.  

Additionally, it is worth mentioning a self-supervised method called contrastive learning. It is a technique to learn the general features of a dataset without labels by telling what data points are similar or dissimilar. The reason why this is important to us is that the similarity of the task is to recognize the similarity. However, the goal might differ for contrastive learning due to their limited dataset or no handcrafted labeling. Several interesting works related to SimCLR, SIMCLRv2, and MoCo (momentum contrast) are fine-tuning with a few labeled examples to achieve high accuracy \cite{chen2020big, he2020momentum, chen2020simple}. These works are essential in unsupervised learning fields and provide highly accurate and efficient solutions. Additionally, an audio similarity work is being introduced to assign high similarity to the audio segment from the same recording while assigning low similarity to different segments from di qfferent recordings \cite{saeed2021contrastive}.

In summary, these systems and applications measure the quality of words, speech, sleep, and emotion. 
However, none of them standardizes the metrics of exercises through the muscular system. PhysiQ recognizes therapeutic exercises and digitalizes the exercises to provide feedback through the metrics and compare exercises using deep learning methods.

\section{Solutions}
\label{sec:solutions}

\begin{figure}%
    \centering
    \subfloat[\centering Exercise Qualitative Feedback on Patients' GUI]
    {{\includegraphics[scale=.085]{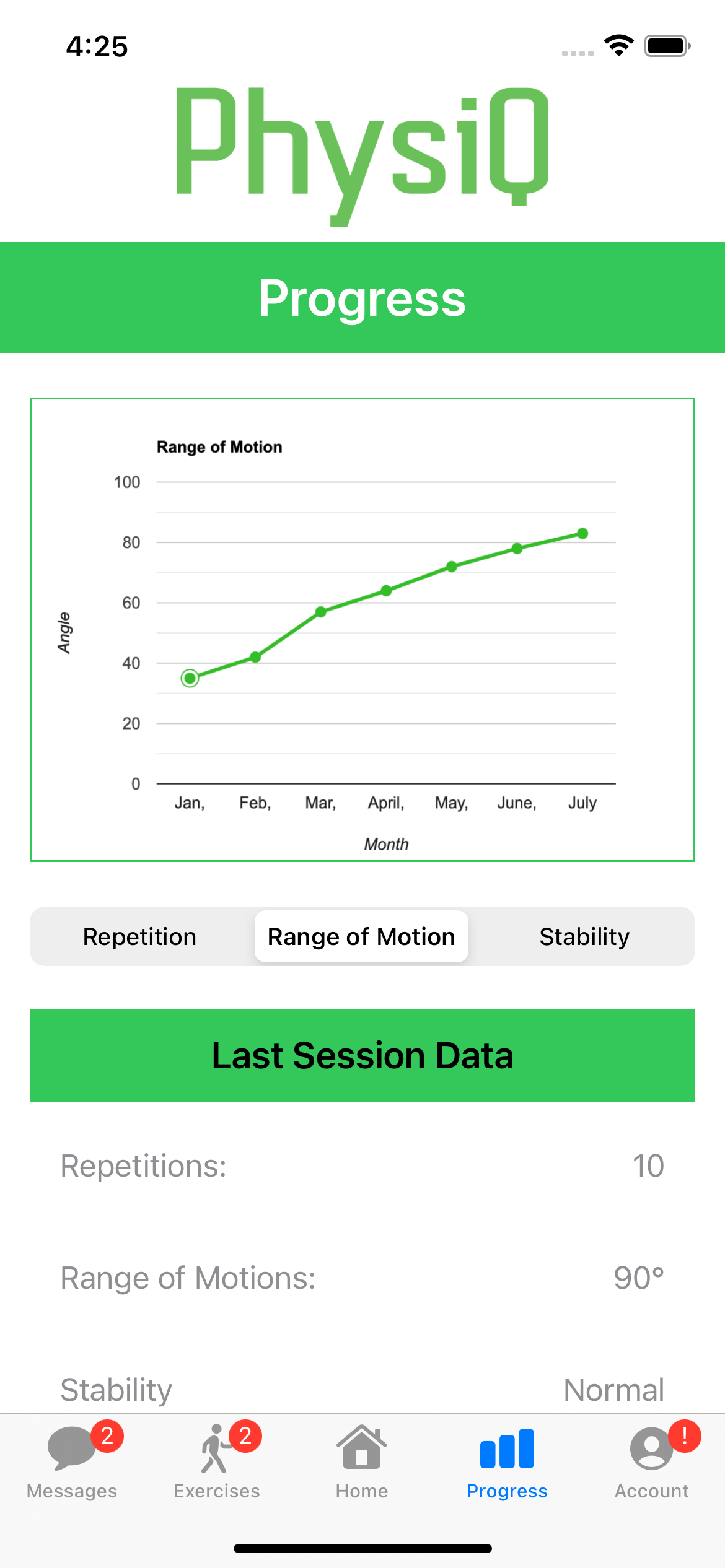} }}%
    \qquad
    \subfloat[\centering TO-DO list on Patients' GUI]
    {{\includegraphics[scale=.085]{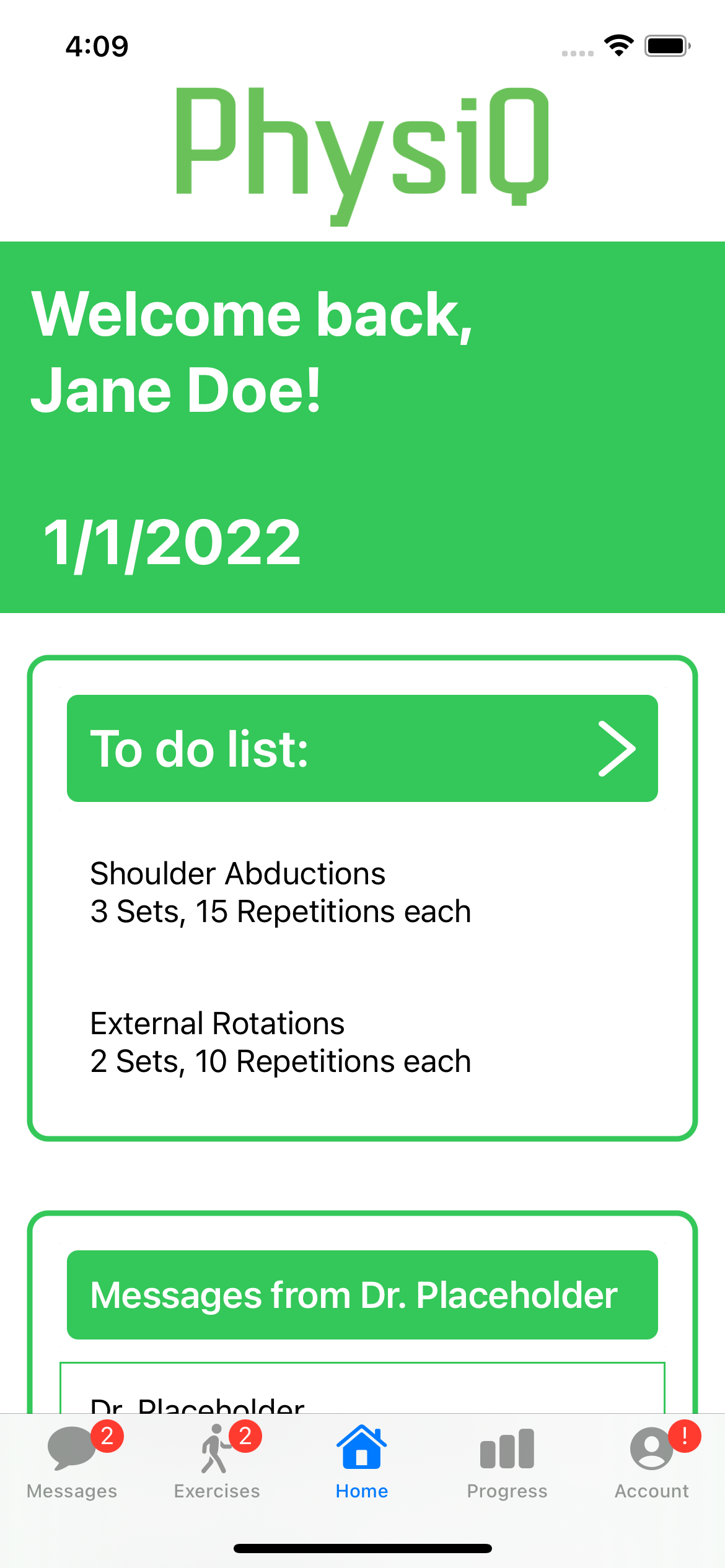} }}%
    \qquad
    \subfloat[\centering Doctors' GUI on Patients' tasks]
    {{\includegraphics[scale=.085]{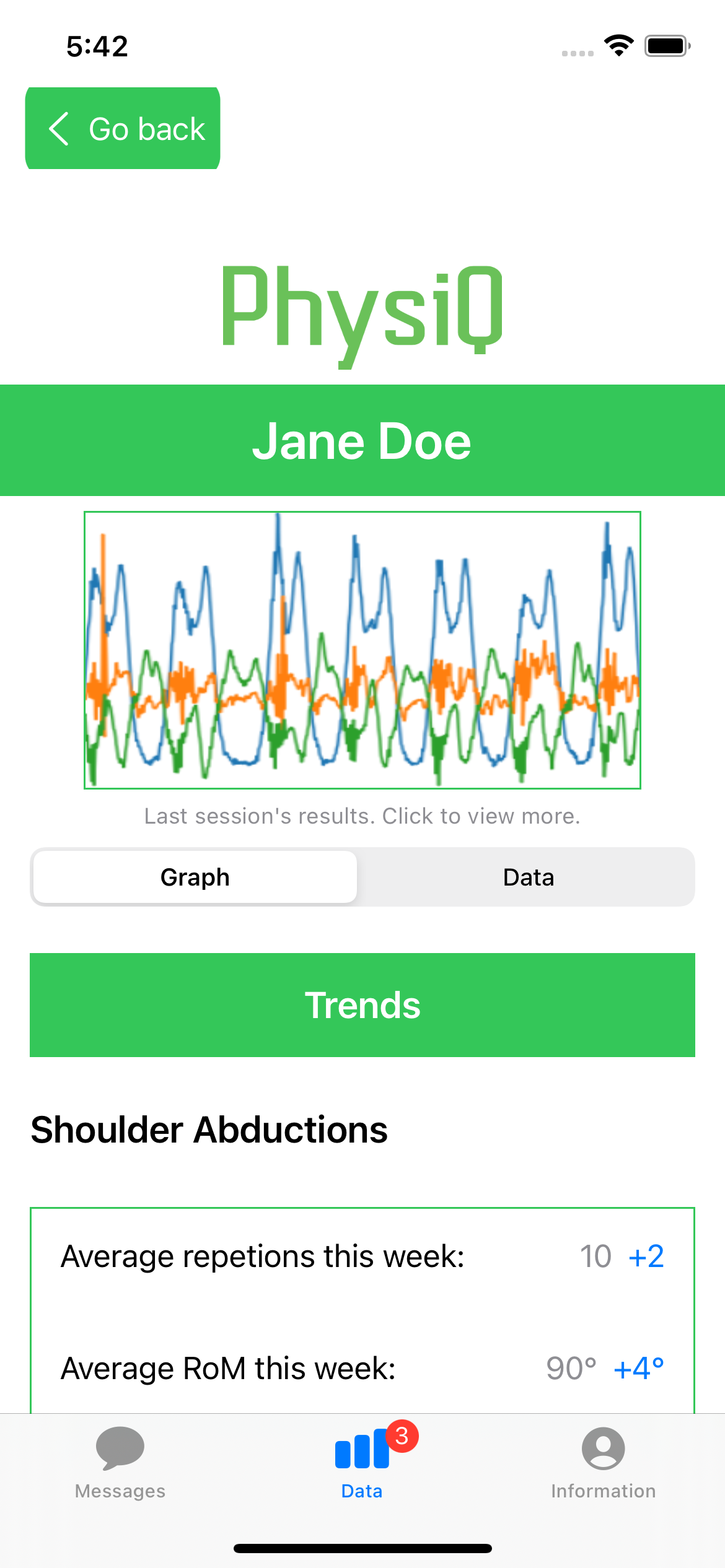} }}%
    \caption{PhysiQ on Smartphone Graphical User Interface (GUI) for patients and doctors. Leftmost GUI demonstrates the patients' progress tab and how they make progress throughout the months; rightmost GUI demonstrates the doctor' tab to monitor patients' results and how much they improve or aggravate with visualization of the data.}%
     \label{fig:PhysiQ_GUI}%
\end{figure}

\begin{figure}[t]
    \centering
    \includegraphics[scale=.5]{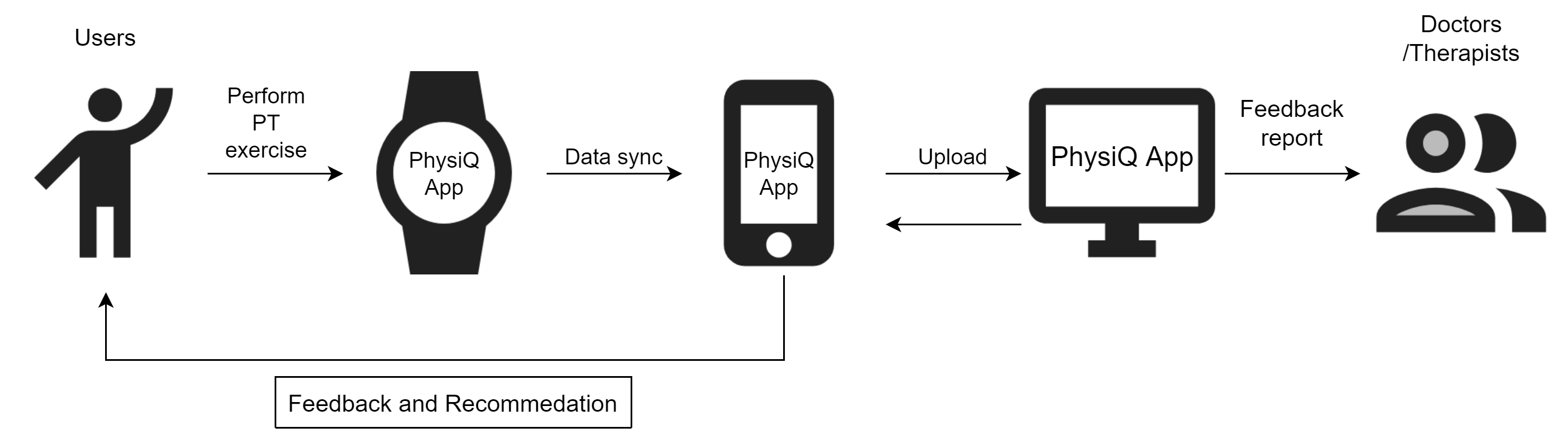}
    \caption{Overview of PhysiQ framework}
    \label{fig:SystemStructure}
\end{figure}

We build PhysiQ to continuously monitor users' off-site exercise activity and quantitatively measure the quality. We particularly measure the quality of exercises in three metrics of \textit{range of motion}, \textit{stability}, and \textit{repetition}. The framework is shown in Fig. \ref{fig:SystemStructure}. 
PhysiQ apps run on three platforms. 
Users first perform activities wearing a smartwatch. The PhysiQ app on the smartwatch extracts the sensory data and syncs the data with the smartphone and cloud in real time. Then, the model on the cloud generates scores for the exercises in different manners, such as range of motion and stability. Based on the scores, the model sends feedback, such as, ``The score for a range of motion of this exercise is 150, which means you did a good job on range of motion. '' and recommendations, such as ``You could do 3 more repetitions!'' on the PhysiQ app on user's smartphone. Meanwhile, it also uploads the user's progress to the cloud for the therapist to review. We present some of the graphical user interfaces (GUI) of our app on the phone in Figure \ref{fig:PhysiQ_GUI}. In the rest of this section, we first formalize our problem, then present how we digitalize the exercise metrics on sensory data, and finally elaborate on the details of our multi-task spatio-temporal SNN-based quality measurement model. 

\subsection{Problem Formulation}\label{subsec:solutions:problem} 
We formalize the problem as follows: given the smartwatch with built-in IMU sensors, it returns the input of IMU sensory data $\mathbb{X}^j_i$ by $j$-th participant and $i$-th sample. Each data $\mathbb{X}^j$ contains $T^j$ number of samples \{$x^j_1$, ..., $x^j_{T^{j}}$\}. Each $x^j_i$ is comprised of 3-axis of accelerometer data ($A^x, A^y, A^z$) and 3-axis of gyroscope data ($G^x, G^y, G^z$). Our model outputs a similarity score $s$.

We further divide our problem into three folds for three metrics: 


(1) Our problem input is $x^j_i$ and output is a set of $X' = \{x'\}$ with $x'$ represents one repetition and $n$ represents total repetitions.

(2) In \textit{range of motion} metrics, we divide it into an absolute and relative value. First, we formalize our problem under the same problem of input sensory data, $x' \in \mathbb{X}^j_i$. In absolute value, for each $x'$, we define $y_{rom}(x')$ as the absolute score of \textit{range of motion}. Secondly, in relative value, we define the relative value, $s_{rom}(x'_i, x'_{i*})$, as similarity score under \textit{range of motion} metrics, where $i*$ is the anchor exercise's index.

(3) For $x' \in \mathbb{X}^j_i$, We define the relative value, $s_{stability}(x'_i, x'_{i*})$, as similarity score of \textit{stability} metrics.

\subsection{Digitalizing Exercise Metrics and Ground Truth} 
\label{subsec:solution:digitalizingExerMetrics}
\label{subsec:data_collection:gt}



\subsubsection{Repetition}
\begin{figure}%
    \centering
    \subfloat[\centering Energy Plot with Accelerometer Data]
    {{\includegraphics[width=7cm, height=5cm]{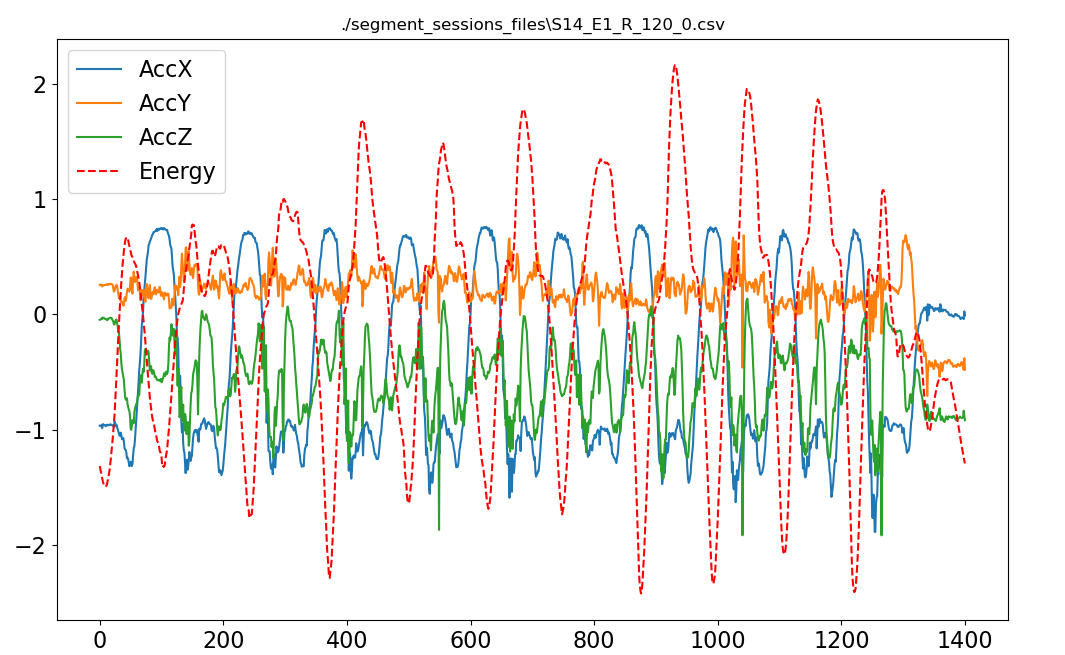} }\label{fig:energy:a}}%
    \qquad
    \subfloat[\centering Potential Cutting Position with Energy Plot]
    {{\includegraphics[width=7cm, height=5cm]{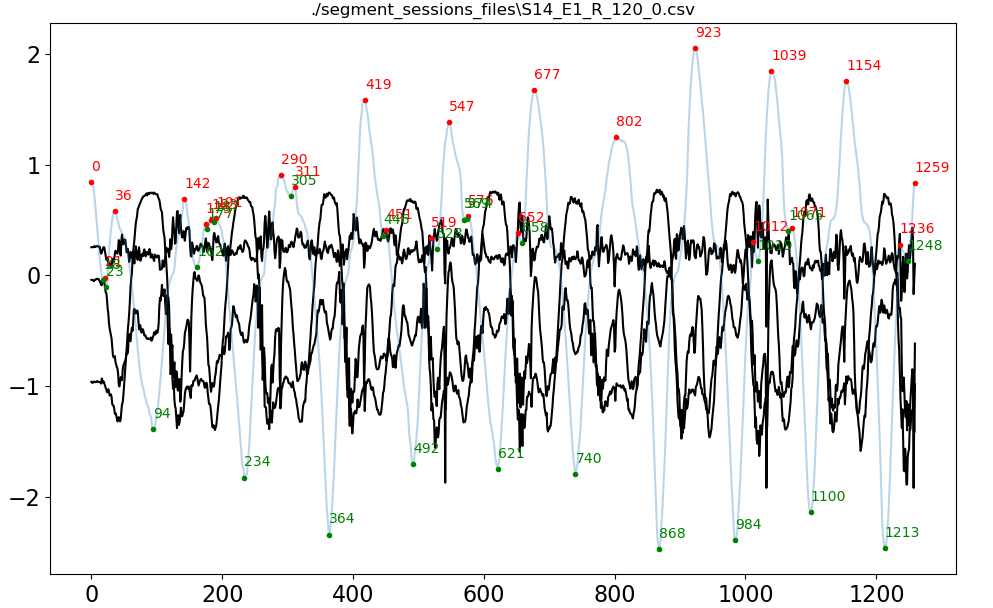} }\label{fig:energy:b}}%
    \caption{These two figures show how the energy plot suggests the cutting positions for the repetition of 10 exercises. For Fig. \ref{fig:energy:a}, the energy is plotted in a red dashed line. The specialist manually does the exercises' beginning and ending cut-off procedure. In Fig. \ref{fig:energy:b}, we change the original data colors to black and emphasize the color of the cutting position for segmentation and energy plot for visualization purposes.}%
    \label{fig:energy}%
\end{figure}

We implement a novel energy function to segment the exercise's repetitions. We calculate the energy as shown below:
\begin{equation} E(i) = \frac{1}{f_s+1}\left(h(i) + \sum_{n=-T}^{T} \sqrt{h(n+i)}\right),  \mbox{where} \: f_s = 50 Hz, T = \frac{f_s \lambda N}{2000} 
\end{equation}

\begin{equation} h(i) = |A^x(i) * W^x| + |A^y(i)* W^y| + |A^z(i)* W^z|
\end{equation}
In the formula above, $i$ is the positional index to calculate the energy throughout the signals, and $N$ is the actual length of the signals in 10 repetitions. As shown in Fig. \ref{fig:energy}, we use this method of calculation to find the cutting position to semi-automatically segment the signal of 10 repetitions to the number of 1 repetition. Noted, since some exercises have a low and high amplitude signal, we use hyper-parameters $W^x, W^y, W^z$, and $\lambda$ weights to adjust the smoothness of the energy. The purpose of the energy is to merge two peaks from the previous and next repetition to form a significant energy level to identify the cutting position. As a result, good hyper-parameters are required to pre-process the segmentation of the data. 

\begin{figure}[t]
    \centering
    \includegraphics[scale=.5]{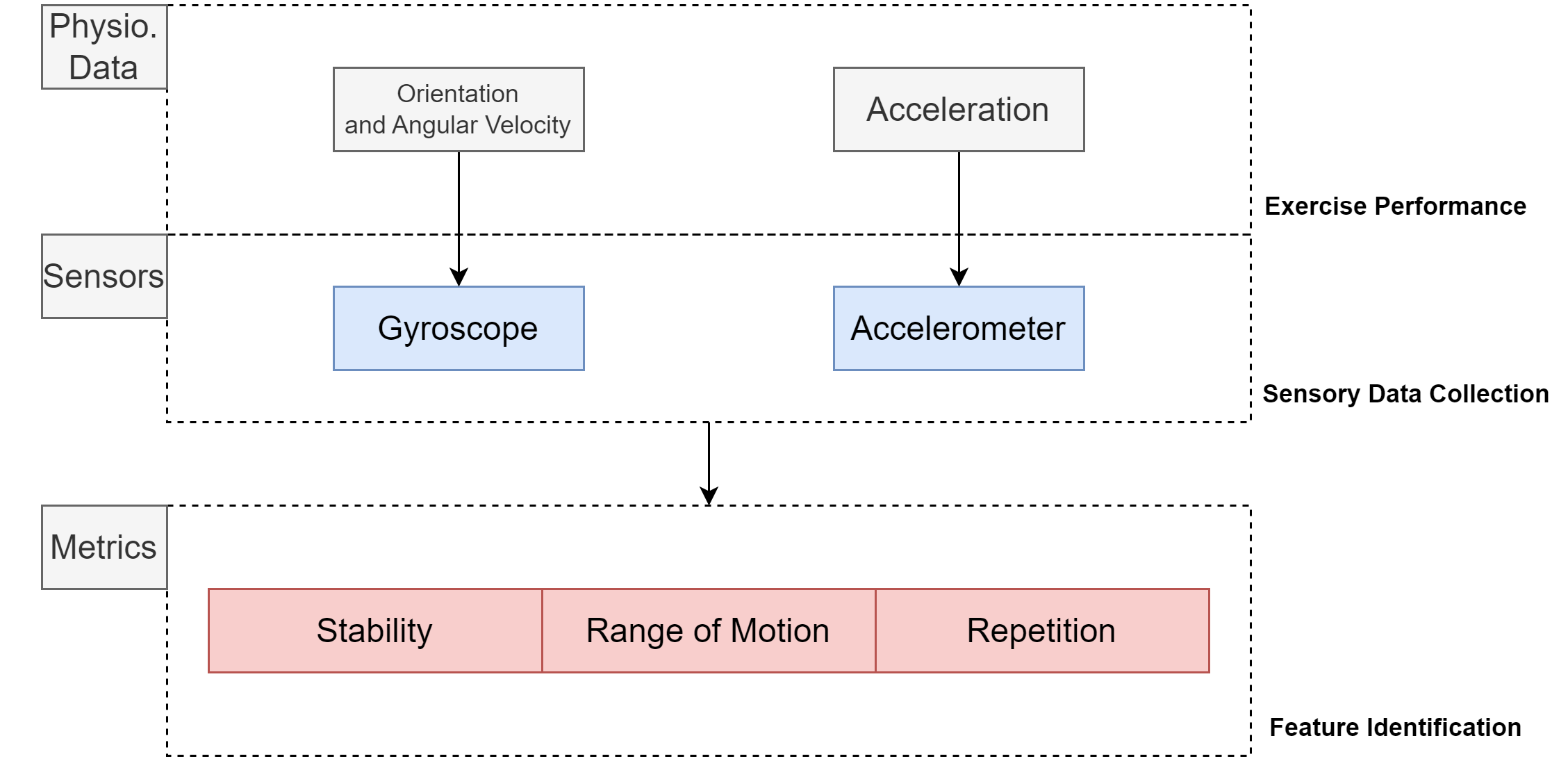}
    \caption{Exercises Metrics: we first identify the type of exercise to perform and which sensory data is collected. Based on the signal data gathered from participants, we measure them against our metrics. The framework assesses the quality of exercises based on the \textit{range of motion}, \textit{stability}, and \textit{repetition}.}
    \label{fig:ExerciseMetrics}
\end{figure}

\subsubsection{Range of Motion}
After using the energy method to segment the signal of participants' exercises, we annotate each exercise according to its labels and positions. In our case, we have three potential labels: \textit{range of motion}, \textit{stability}, and \textit{repetition}. We modify our method to generate \textit{stability} using Equation \ref{eq:gt-stb} as our ground truth for stability. \textit{Range of motion} metrics are collected through participants' exercises under the supervision of experimenters. We examine the participants' range of motion as they performed and verify with recorded videos. \textit{Repetition} is labeled based on the number of repetitions merged together. We target one particular exercise to build our framework and use two additional exercises to verify its competency: shoulder abduction (SA), external rotation (ER), and forward flexion (FF). We explain the process in Section \ref{sec:evaluation}. Due to different muscle activation of \textit{range of motion}, we examine shoulder abduction as our initial exercise to design our metrics and framework. Additionally, we classify shoulder exercises into two main categories: half arm span (HAS) and half-half arm span (HHAS). HAS contains exercises that require both forearm and arm into motion. HHAS only requires forearm or arm into motion.

\begin{figure}
    \centering
    \subfloat[\centering Shoulder anatomy, front view]
    {{\includegraphics[width=5cm]{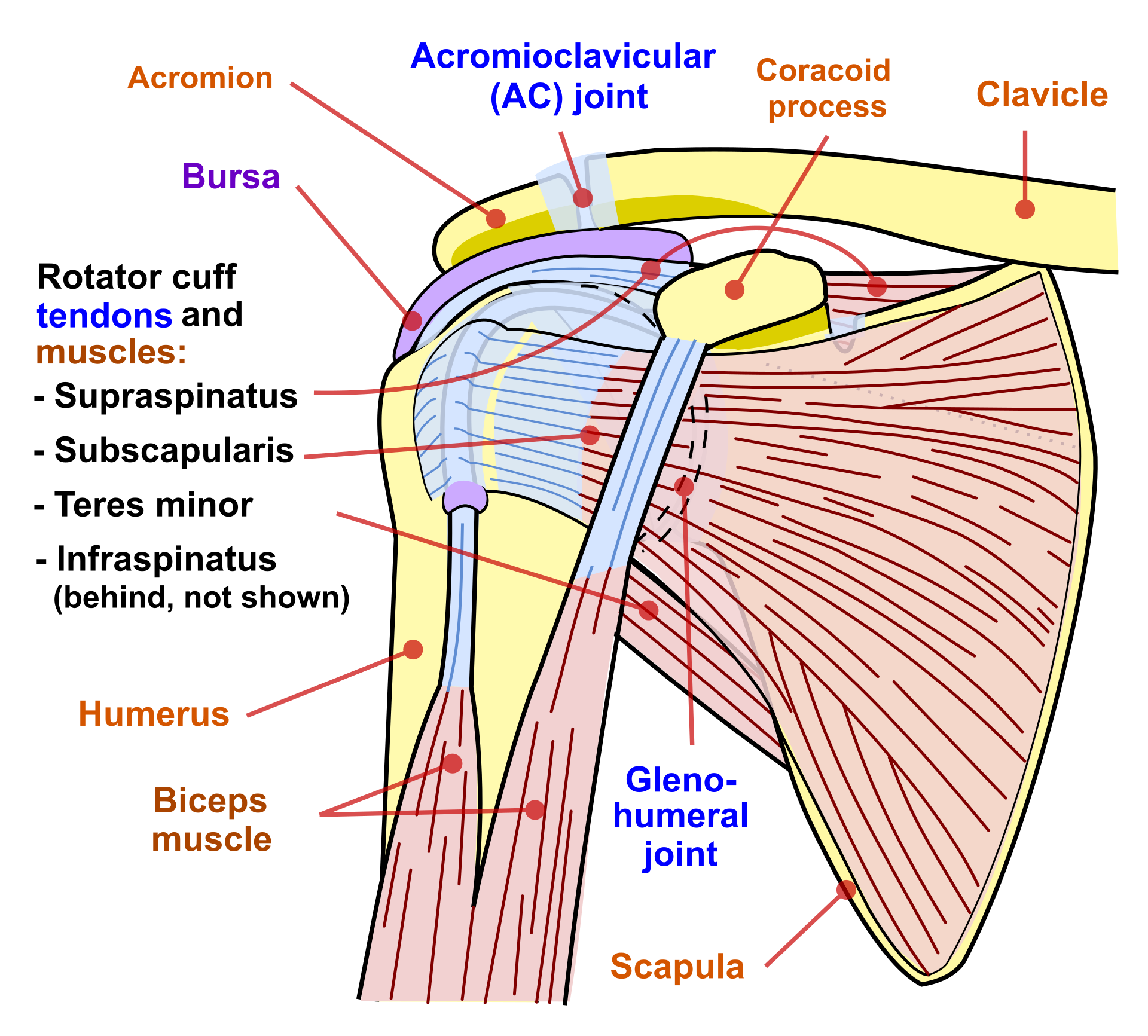} }}%
    \qquad
    \subfloat[\centering Serratus anterior muscle]
    {{\includegraphics[width=5cm]{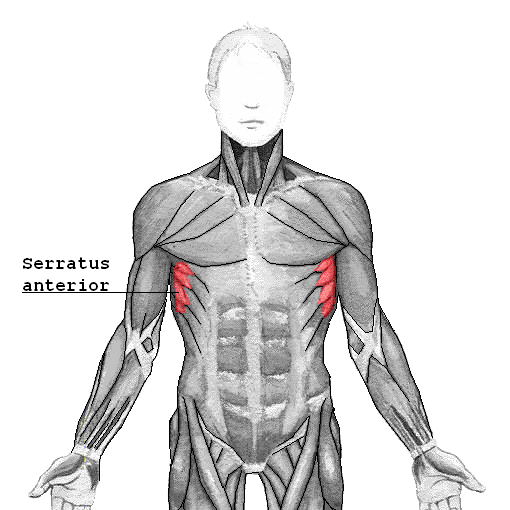} }}%
    \qquad
    \subfloat[\centering Deltoid muscle]
    {{\includegraphics[width=5cm]{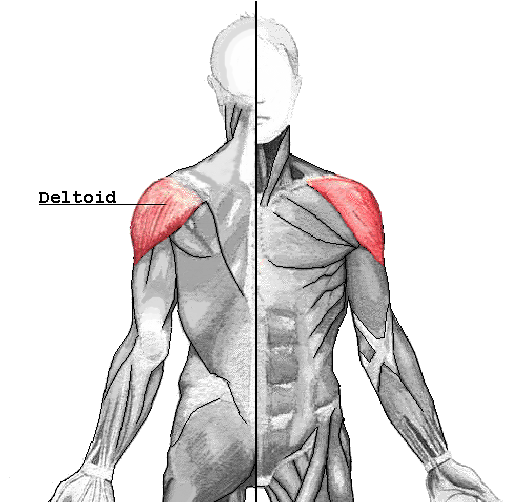} }}%
    \qquad
    \subfloat[\centering Trapezius muscle]
    {{\includegraphics[width=5cm]{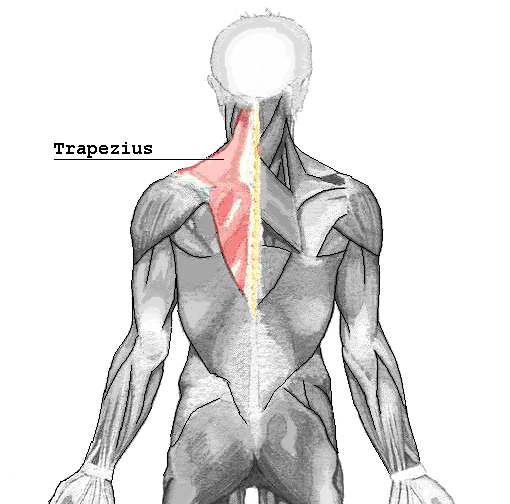} }}%
    \caption{This is a visualization of how shoulder with joints, tendons, and muscles work in a close look \cite{muscle:shoulder, muscle:deltoid, muscle:serratusanterior, muscle:trapezius}.}
    \label{fig:shoulderjoint}
\end{figure}

In shoulder abduction exercise, it involves the glenohumeral joint and scapulothoracic articulation in different \textit{range of motion}. At first 20 to 30 degrees of motion, subjects do not use the scapulothoracic joint motion, and the supraspinatus tendon should be the only muscle helping during this. Deltoid muscles are activated to support from 30 to 120 degrees of range of motion. Lastly, beyond 120 degrees, a full abduction is considered when the arm is externally rotated with the humerus activating. Different range with different muscles activation inspires us to finalize five different \textit{range of motion} and \textit{stability} (we mentioned on how to create stability in Section \ref{sec:data collection}) as our categories to understand the quality of exercises with Fig. \ref{fig:shoulderjoint} \cite{muscle:shoulder, muscle:deltoid, muscle:serratusanterior, muscle:trapezius}.

\begin{figure}[hbt!]
    \centering
    \includegraphics[scale=.5]{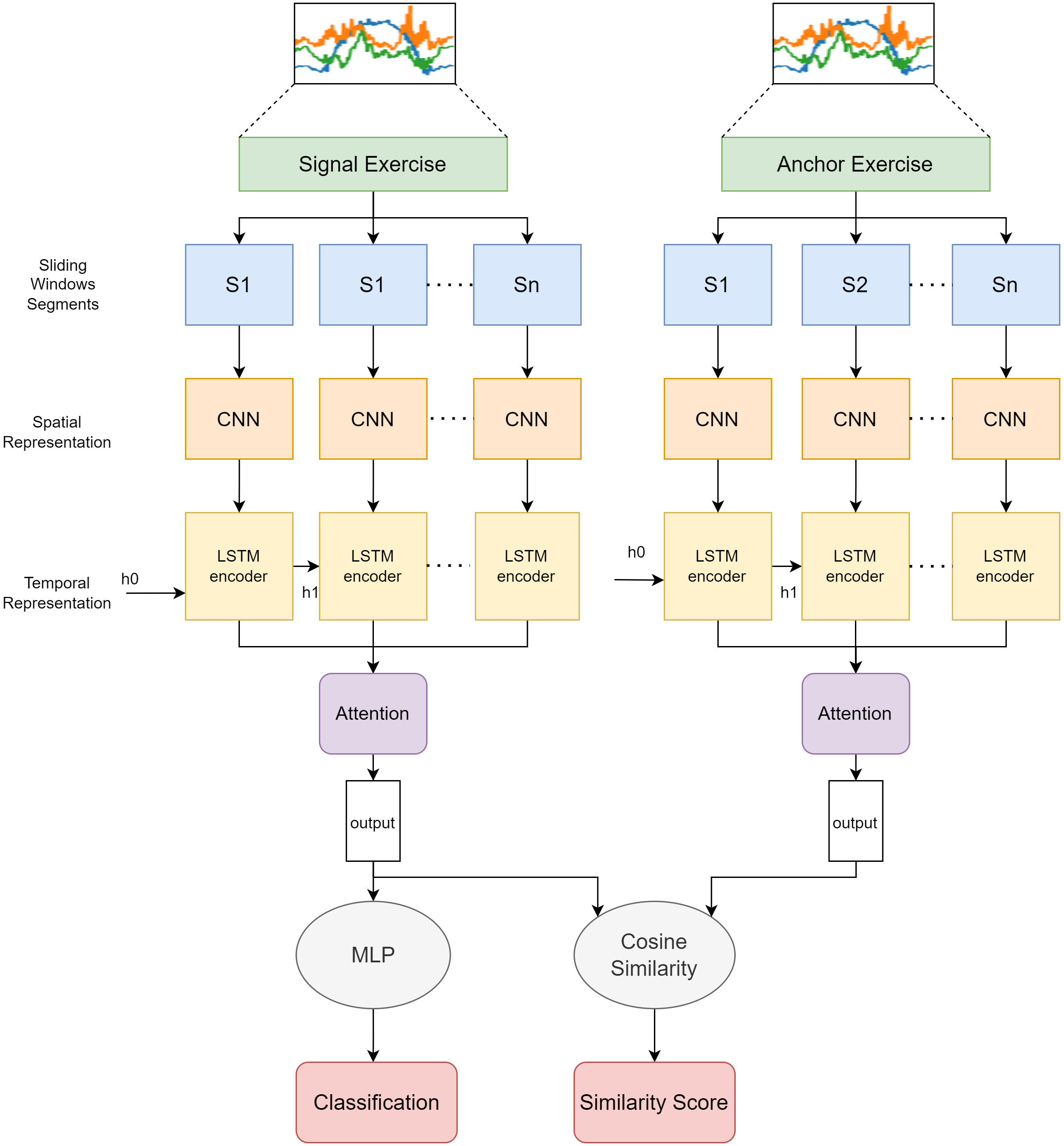}
    \caption{Multi-task Spatio-temporal SNN: First of all, we have 2 one repetition exercises feed into our network as signal and anchor exercise. Signal exercise can be compared against anchor and vice versa. After Sliding Windows Segmentation, Spatial and Temporal encoding, we feed our model into an attention mechanism. Finally, we compare these two hidden features representation and output its similarity score using cosine similarity. Additionally, the hidden feature of the signal exercise is processed into a MLP network to get result of classification based on the metrics of \textit{range of motion}, \textit{stability}, or \textit{repetition}.}
    \label{fig:SiameseNetwork}
\end{figure}

\noindent\textit{Range of Motion for HAS Exercise Muscular Activation }
\begin{itemize}
    \item 30 degree ROM: supraspinatus muscles 
    \item 60 degree ROM: deltoid muscles
    \item 90 degree ROM: deltoid muscles, transition to trapezius muscle
    \item 120 degree ROM: trapezius and serratus anterior muscle
    \item 150 degree ROM: trapezius, serratus anterior muscle, and humerus activation
\end{itemize}

In external rotation exercise, the rotator cuff is used to perform this exercise. The rotator cuff consists of four muscles that stabilize the shoulder in external rotation, infraspinatus, teres minor, supraspinatus, and subscapularis as shown in Fig. \ref{fig:shoulderjoint}. As in external rotation, the infraspinatus muscle stabilizes the shoulder joint and acts as the prime mover in this exercise \cite{jang2014changes}.

\noindent\textit{Range of Motion for HHAS Exercise Muscular Activation}
\begin{itemize}
    \item 45 degree ROM: minimally infraspinatus muscle
    \item 90 degree ROM: infraspinatus muscle, posterior deltoid
    \item 150 degree ROM: infraspinatus muscle, posterior deltoid, and all other muscles in rotator cuff.
\end{itemize}

Similarly, in forward flexion exercise, the muscle activation includes supraspinatus, infraspinatus, and anterior deltoid. We classify \textit{range of motion} into 5 similarly to shoulder abduction because these two exercises are considered half arm span exercise.

\subsubsection{Stability}
Initially, we defined our ground truth of stability using resistance bands relative to the participants' strength. For example, if the participant is strong, we progressively find his or her maximum strength with the resistance band and create two different instability based on that, labeled as two classes. However, this method does not guarantee that stability correlates with the levels of resistance bands. Furthermore, as we visualize and evaluate it, we do not find the difference between the two different resistance bands. Therefore, we define the stability metrics using a low pass filter and coefficient of variation through a mathematical methodology. By doing so, we can measure stability of all the exercises performed by participants. Low pass filter is used to differentiate what is human motion and signal noise since our IMU sensors captures at 50 Hz. As suggested by Khusainov et al., human activity frequencies are between 0 and 20 Hz, and 98\% are below 10 Hz \cite{khusainov2013real}. Therefore, for maximal capturing of stability, we use 20 Hz as our parameter for the low pass filter function.

\begin{equation} \label{eq:gt-stb}
instability(S) = Tanh\left(|\:CV(\mbox{lps}(S, \:\mbox{f=20}))\:|\right), \mbox{where} \: S = [A^x, A^y, A^z, G^x, G^y, G^z]\mbox{.T} \end{equation}
\begin{equation} \label{eq:gt-stb-sub1}
CV(S) = \frac{\mu_S}{\sigma_S} \end{equation}

In the formula above, the lps represents the trivial low pass filter function we used to filter some noises.

\noindent\textit{Stability Muscular Activation}
\begin{itemize}
    \item 0.0 stability: stable and perform normal exercises
    \item in-between 0.0-1.0 stability: rotator cuff
    \item 1.0 stability: (unstable) supraspinatus, infraspinatus, teres minor, and subscapularis 
\end{itemize}


\subsection{Spatio-temporal Feature Representation}


As shown in Fig. \ref{fig:SiameseNetwork}, we develop a method of combining spatial and temporal representation to recognize the shape of signals of exercises \cite{murahari2018attention, chen2019multi, bhattacharya2020step, peng2018aroma}. We use the attention mechanism as a method of message passing to understand relationships in different hidden features of sliding window segments in one repetition. 

To discern and recognize the details of the signals and find the correlation between metrics and exercises, we tackle the problem in two different aspects: time and space. The exercises performed by users are the input of temporal signals. The signals result in a pattern in space to depict different angles of the exercises. 

We build a CNN-based spatial encoder as: 

\begin{equation}
    CNN(w) = (\sigma(sum(W_1 \odot  w) + b_1), \ldots, \sigma(sum(W_n \odot w ) + b_n)),
\end{equation}
where $W_1, W_2, \ldots, W_n$ are learnable weights matrices, $b_1, b_2, \ldots, b_n$ are biases, $\odot$ is element-wise multiplication, $sum$ is element-wise summation, and $\sigma$ is the activation function such as ReLu. CNN is capable of effectively interpreting spatial information and transforming it into a hidden pattern. It has the potential to compress information to represent into a smaller space, and it is very effective to compress sliding windows of signals. This provides a feature extraction mechanism across windows with specific weight matrices and biases. What is important in spatial encoding is by using a feature extraction method, our model can interpret the importance of each window given its features. 
Additionally, a max pooling is applied in our CNN model to reduce the signal's dimension.

However, understanding the features in each window is not enough to distinguish any temporal knowledge due to the property of time series, such as trends, and seasonal or non-seasonal cycles. Next, we employ Long Short Term Memory networks (LSTM) for temporal encoding as \cite{hochreiter1997long}:

\begin{equation}
    f_t = \sigma(W_f  x_t + U_f h_{t-1} + b_f)
\end{equation}
\begin{equation}
    i_t = \sigma(W_i  x_t + U_i h_{t-1} + b_i)
\end{equation}
\begin{equation}
    o_t = \sigma(W_o x_t + U_o h_{t-1} + b_o)
\end{equation}
\begin{equation}
    \tilde{c_t} = tanh(W_c x_t + U_c h_{t-1} + b_c)
\end{equation}
\begin{equation}
    c_t = f_t c_{t-1} + i_t \tilde{c_t}
\end{equation}
\begin{equation}
    h^r_t = o_t tanh(c_t),
\end{equation}

where, $U$ $W$, and $b$ are weights and biases that is not time dependent, $\sigma$ is a sigmoid activation function, and $i$,$f$,$o$ are input, forget, and output gates respectively. Lastly, $c$ and $h$ are the cell and hidden states vector given the time $t$.

Next, attention mechanism on a set of queries, keys of dimension $d_k$, and values are calculated using the matrix of output as shown below:

\begin{equation}
    Attention(Q, K, V) = softmax\left(\frac{QK^T}{\sqrt{d_k}}\right)V
\end{equation}

Instead of performing one attention function, we apply multi-head attention where head is the number of paralleled attention mechanism. Multi-head attention allows our model to attend information with different representation at different locations in time and space. We use $H$ as the number of head in multi-head attention:

\begin{equation}
    MultiHead(Q, K, V) = concat(head_1,\ldots, head_h)W^O
\end{equation}
\begin{equation}
    head_i = Attention (QW_i^Q, KW_i^K, VW_i^V)
\end{equation}

In our work, we use $H$ = 16 heads of paralleled attention layers. Additionally, we use $d_{model}$ = 256; therefore, each $d_k = d_v = d_{model}/H = 16$.

As a result, we describe our model, PhysiQ. It takes advantage of the fact that every sliding windows $w$ of size $k$ x 6 can be interpreted as a frame in time. Using this, we feed each window $w$ into the CNN to output $h^c$ as $z^c$ x 1. Additionally, we feed $h^c$ into LSTM to get $h^r$ with a size of $z^r$ x 1. The LSTM creates a sequence of hidden states $[h^r_0, \ldots, h^r_n]$, acted similar to a positional encoding to understand the temporal information. We pass the sequence into the attention mechanism returning our hidden representation, $A$ that has both temporal spatial information, and relational message passing knowledge. 

\begin{algorithm}
	\caption{PhysiQ Framework Encoder} 
	\begin{algorithmic}[1]
	\STATE {Def: $A = ENCODER(e)$, s.t. exercise segment $e$ passes in, return $A$, where $A$ has $n$ number of hidden presentation as $h^r_0, \ldots, h^r_{n-1}$}
    \STATE \textbf{Input}: $e$, exercise segment
    \STATE \textbf{Output}: $A$, hidden representation of one exercise segment
        \STATE ${w_0, w_1, \ldots, w_{n-1}}= SLIDE(e)$    \\ $SLIDE$ takes an exercise $e$ and return $W$, where $W$ is the size of $n$ x $k$ x 6, $n$ number of sliding windows.
        \STATE $W = {w_0, w_1, \ldots, w_{n-1}}$
	    \FOR {each $w_i$ in $W$}
	        
	        \STATE $h^c_i = CNN(w_i)$
	        \STATE $h^r_i = LSTM(h^c_i)$
	    \ENDFOR
	    \STATE $H=[h^r_0,h^r_1, \ldots,h^r_{n-1}]$
        \STATE $A = Attention(H, H, H)$
        \STATE return $A$

	\end{algorithmic} 
	\label{algo:encoder}
\end{algorithm}

\begin{algorithm}
	\caption{PhysiQ Similarity Comparison} 
	\begin{algorithmic}[1]
	\STATE {Def: $s_{ij} = SIMILARITY(e_i, e_j)$, where $s$ is the similarity score and $e$ is the signal segment}
    \STATE \textbf{Input}: $e_i, e_j$, a pair of exercise segments
    \STATE \textbf{Output}: $s_{ij}$ is the similarity score between a pair of exercises 
    
    \STATE $A_i = ENCODER(e_i)$
    \STATE $A_j = ENCODER(e_j)$
    \STATE $ s_{ij} = Cosine(A_i, A_j) = A_i \cdot A_j / ||A_i|| * ||A_j||$ 
    \STATE return $s_{ij}$ 
	\end{algorithmic} 
\end{algorithm}






\subsection{Similarity Comparison}
The supervised learning framework has recently improved dramatically on 1-D and 2-D healthcare signal processing tasks. However, it does not leverage the framework to understand the relationship between the inputs. Significantly, how can patients improve without an anchor comparison from the previous performance? By comparing their day-to-day performance, PhysiQ understands whether participants enhance their performance based on the result of their exercises and how the patients are improving. To address this issue, we utilize the Siamese Neural Network, a type of contrastive learning framework that can extract useful features from data itself without the need for large handcrafted labels. On top of that, we design a data collection strategy to gather multi-modality of data for future evaluation analysis.

Siamese Neural Network (SNN) is a neural network that shares and contains two identical networks with the same configuration and the sharing of the weights. The identical model is used to find the similarity between two inputs. At the same time, the advantages of the SNN are more robust to the class imbalance in the data, learning a tremendous hidden and embedding deeply semantic similarity; however, it does not necessarily output probabilities but the distance between classes. We re-design the network and make it to fit the problem of reference comparison in Fig. \ref{fig:ExerciseMetrics}. We formulate a regressive distance between two exercises as the ground truth label to train SNN with a prior assumption of a maximum of the \textit{range of motion} of $R$ in shoulder abduction, as shown below:

\begin{equation}\label{eq:rom}
s_{rom}(m_a, m_b) = 1 - |\frac{m_a}{R} - \frac{m_b}{R}| \end{equation}

With the Equation \ref{eq:gt-stb} to get the signal's \textit{stability}, we can measure the similarity:

\begin{equation} \label{eq:stb}
s_{stability}(S_a, S_b) = 1 - |instability(S_b) - instability(S_b)| \end{equation}

Lastly, with the assumption of maximum of repetition $M$, we can measure the similarity of \textit{repetition}:
\begin{equation}
    s_{repetition}(r_a, r_b) = 1 - |\frac{r_a}{M} - \frac{r_b}{M}|,
\end{equation}
where $r_a$, $r_b$ represent the number of repetition of signal and anchor exercise.



The SNN is very popular and used to solve various problems in research. However, to accommodate our specific problem, we aim to improve our accuracy by carefully designing our feature extraction in spatial and temporal aspects. Therefore, 
we adapt the SNN for our signal comparison because there is an underlying knowledge and information that the deep learning method interprets and understands. Additionally, we leverage the temporal encoding, spatial encoding, and attention mechanism to generalize the model with our metrics. Additionally, we use cosine similarity as our similarity measurement because of its ability to differentiate orientation distance between two encoded exercise features.

\section{Data Collection}
\label{sec:data collection}

\subsection{Type of Exercises}
\begin{figure}%
    \centering
    \subfloat[\centering Shoulder Abduction]
    {{\includegraphics[scale=.15]{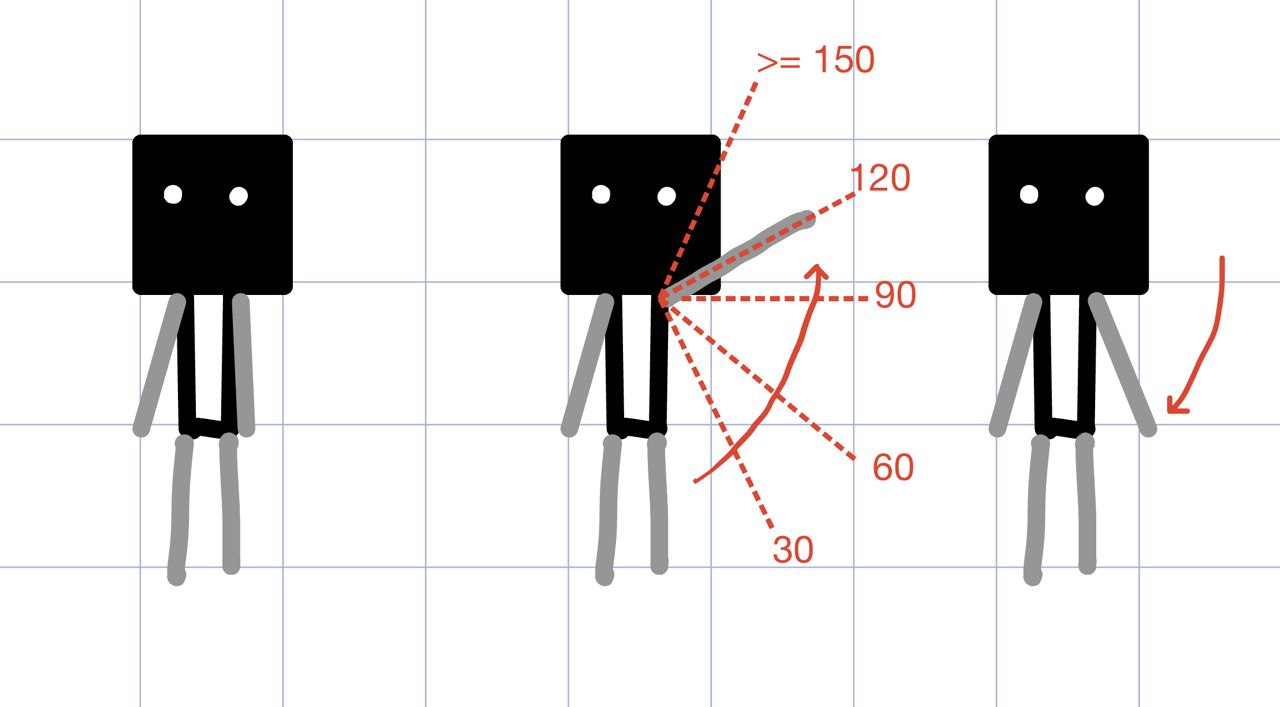} }}%
    \qquad
    \subfloat[\centering External Rotation]
    {{\includegraphics[scale=.15]{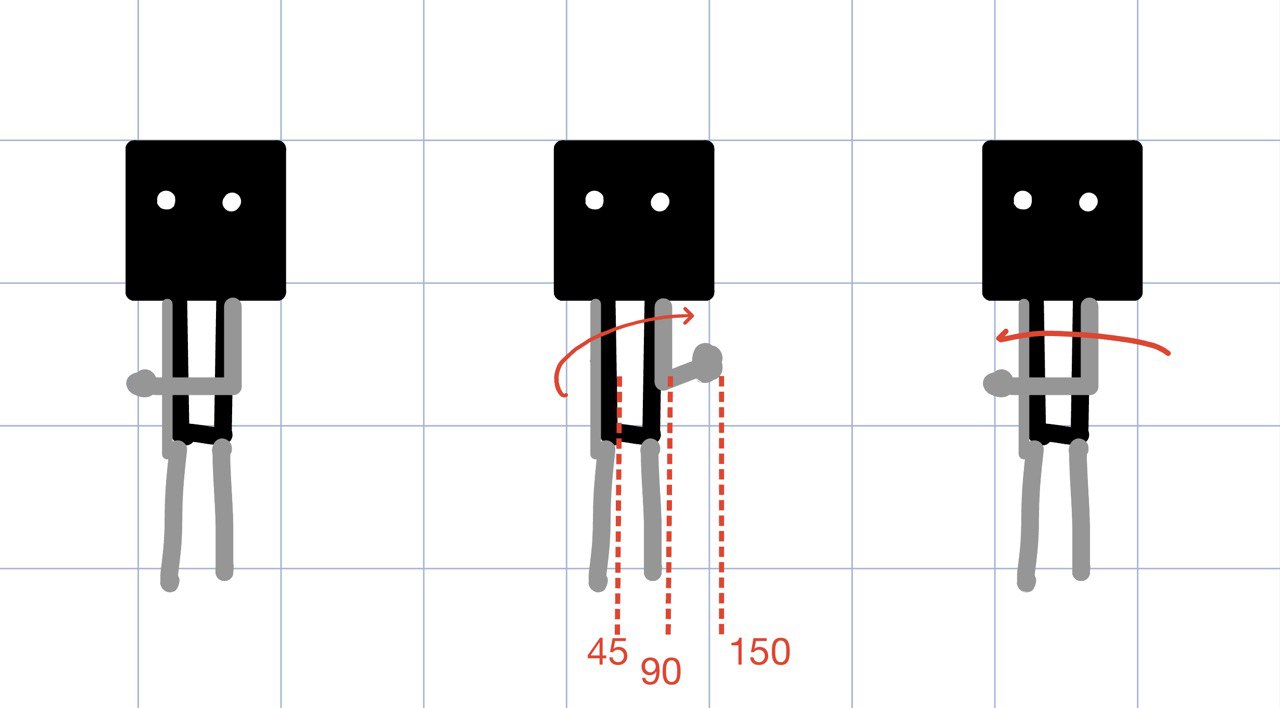} }}%
    \qquad
    \subfloat[\centering Forward Flexion]
    {{\includegraphics[scale=.15]{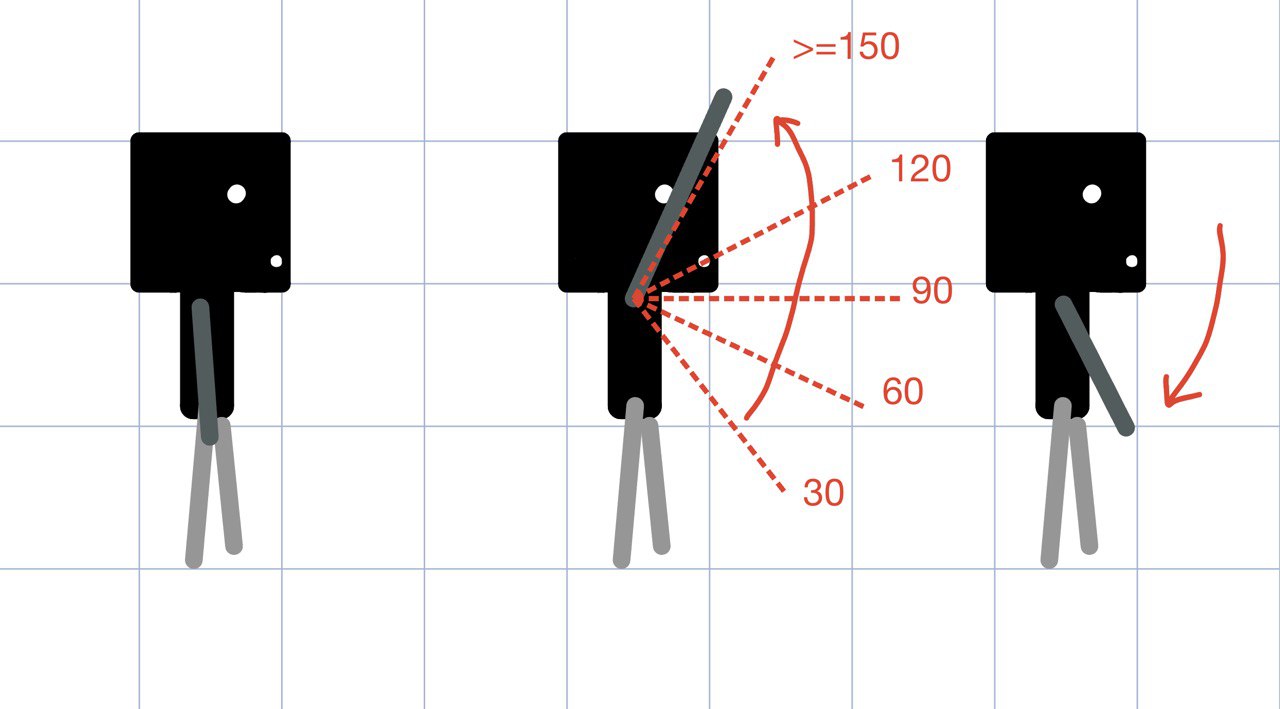} }}%
    \caption{Three Exercises: The exercise on the top left is one repetition of shoulder abduction, the exercise on the top right is one repetition of external rotation, and the exercise on the bottom is forward flexion. Also noted is that we define greater equal than 150 degrees of motion as 150 on the shoulder abduction and forward flexion.}%
    \label{fig:Exercises}%
\end{figure}

Selecting well-represented exercises is very crucial to our problem because the exercise itself should have the quality of repetitiveness, singularity (a defined beginning and ending), reconstructiveness (an exercise that can reshape part of a person's body as an improvement), and representativeness (can cover many parts of the muscles). Therefore, considering these factors, we conduct our evaluation of these three exercises, as shown in Fig. \ref{fig:Exercises}:
\begin{itemize}
    \item Shoulder Abduction: The shoulder abduction is an exercise that requires subjects to stand straight with both hands tucking on the side of the legs as the starting point. Subjects perform the exercise by raising the instructed arm to a certain degree of motion and straightening the arm. Once the subjects reach the stopping point, they drop their arm steadily, as it is similar to raising their arms. 
    \item External Rotation: The external rotation has two components. First, the upper arm (bicep and tricep) tucks in the armpit while rotating the arm externally and keeping the lower arm raised to about 90 degrees, perpendicular to the chest or abdominal. Secondly, the arm should move horizontally and stop at a certain degree of motion or to the full ROM, where the subject's shoulder should feel a sense of blocking by the joints. 
    \item Forward Flexion: The forward flexion is similar to shoulder abduction but different in direction. Subjects perform the exercise by raising the arm slowly forward, reaching the point of ROM, then consistently lowering the arm to the beginning position.
\end{itemize}


We simplify our problem as a preliminary examination from the description we mentioned above. Each exercise has its advantage and weakness. As a result, shoulder abduction, external rotation, and forward flexion have aspects that we look for in exercises, including subjects not needing to lie down and wear a smartwatch on the leg to perform exercises in initial modeling. Additionally, we choose shoulder abduction as our first exercise to test our framework, PhysiQ, because shoulder abduction has all the factors we considered: repeating in sets, singular in the beginning and ending, reconstructing people's bodies, and targeting many muscles around the shoulder. It requires extensive time and resources to find the participants during the pandemic, and the quality of data is varied based on the subjects because we minimize how much we tell the subjects to perform the exercises while giving enough details on reaching different stop points of degrees in motions and instability. In order to perform the metric of \textit{range of motion}, \textit{stability}, and \textit{repetition}, we see how robust our model is and in what scenario it can handle and fail.

\subsection{Data Statistics}
In total, we have 1550 segmented one-repetition exercises of \textit{range of motion (ROM)} and 1170 segmented one-repetition exercises of \textit{stability} for shoulder abduction. At the same time, we have 600 segmented one-repetition exercises of ROMs for external rotation. Additionally, the third exercise, forward flexion, has 650 segmented one-repetition data.
In total, we have collected 31 participants for all data collection in different evaluation periods.
In addition, we have 31 participants from shoulder abduction, 24 participants from external rotation, and 11 participants from forward flexion. The metrics of \textit{range of motion} are labeled as we collect the data, and the \textit{stability} is generated using our method as shown in Equation \ref{eq:gt-stb}. The metric of \textit{repetition} is utilized through our energy segmentation and merged based on the number of repetitions for evaluation. We can form any number of \textit{repetition} by combining the adjacent neighbors of segmented repetitions.

In SNN, we create pairs of inputs for similarity comparison in one repetition exercise of \textit{range of motion}, \textit{stability}, and \textit{repetition}. Therefore, we define the problem as no comparison between subjects, only the comparison of exercises within one particular subject at a time, because with the presumably perfect anchor exercise, it is a relative measure of the similarity of the signal exercise on a particular user.




\begin{table}[h!]
\centering

\begin{tabular}{||c || c | c |c |c||} 
 \hline
  Attribute & Male & Female & avg & std\\ [0.5ex] 
 \hline
 \hline
 Gender & 17 & 14&N.A.&N.A. \\
  \hline
 Age & 18-25 & 18-44 &22.32&4.77\\
  \hline
 Height (cm) &  167.6-190.5     & 149.8-182.9&173.6&10.27 \\
  \hline
 Weight (kg) &  54.4-108.8      & 45.3-88.9&67.31&15.90\\
  \hline
 Previously Shoulder Injures & 2 & 1&N.A.&N.A. \\
 \hline
\end{tabular}

\caption{Participants information}
\label{table:datastats}
\end{table}


\section{Evaluation}
\label{sec:evaluation}


We evaluate PhysiQ from three perspectives: model performance in exercises, generalizability in different metrics, and importance of components in the model. We compare the overall performance of the framework with different types of baseline and how the model performs in different exercises and metrics. We explain the design and set up in Section  \ref{evaluation:subsec:implementation}, \ref{evaluation:subsec:traintest}. Additionally, we test our model performance and generalizability in Section  \ref{evaluation:subsec:performance}. Lastly, we evaluate our model's components in Section  \ref{evaluation:subsec:parameters} and Section  \ref{evaluation:subsec:ablation}. 
\subsection{Implementation}\label{evaluation:subsec:implementation}
We implement the PhysiQ application in consumer-level IOS Apple Watch with an automatically connected application on iPhone. PhysiQ uses a built-in accelerometer and gyroscope to collect sensory motion data and analyze the data quantitatively. The maximum sampling rate that we choose is 50 Hz, because through our analysis of a single exercise, we observed that the Fourier frequency is mostly below 10 Hz, and other papers also support this observation \cite{HUMAN_MOTION_FREQUENCY_2013}. Additionally, one of our metrics is stability, and such core motion of the body should be captured more cautiously with a higher frequency rate. Thus, 50 Hz is what we decide to use. Lastly, once the users have performed exercises, the result automatically synced from the watch to the smartphone through Xcode WCSessions. 

\subsection{Training and Testing Dataset}\label{evaluation:subsec:traintest}
Once the data is recorded, we segment the data according to our energy function. We used weights $W_x$,$W_y$,$W_z$ for each accelerometer x, y, and z, and $\lambda$ as an additional hyper-parameter. Next, we split the dataset and apply standard scales for all exercise segments; the scaler is an axis-wise scaler standardized on the current axes (x, y, and z for both accelerometer and gyroscope sensory data). There are two splitting methods for evaluation we employed. The first one is leave-one-person-out cross-validation (LOOCV). LOOCV is a cross-validation approach that treats each subject as a “test” set. This type of k-fold cross-validation has the k value as the number of participants. LOOCV separates the models from seeing the validation/testing set. As a result, the model does not see the distribution of validation subjects (participants), and we can analyze the generalizability of our model on 34,000 training samples and about 3,000 validating and testing samples, of total 31 participants. In ROM and stability, we have a total of 37000 data samples. In repetition, we perform a repetition comparison among 1, 2, and 3 repetitions, with a total of 280,000 data samples in shoulder abduction, 15,000 in external rotation, and 8,000 in forward flexion. For efficiency, we randomly choose 10\% of the data in shoulder abduction for the overall evaluation 10 times. Secondly, we perform a standard 70\% 10\%, and 20\% respectively on training, validation, and testing splits as our overall evaluation. We randomly extract these splits in each subject in 70, 10, and 20 fashions in normal splits. By having both evaluations, we should know how our model works in real-world scenarios and how well our model can perform. 


\subsection{Evaluation Metrics}\label{evaluation:subsec:evaluationmetrics}
We evaluate the performance using three measure methods in all experiments, i.e., R-squared, Mean Square Error (MSE), and Mean Absolute Error (MAE). R-squared, the coefficient of determination, is how close the data are fitted in the model or the percent of variation explained by the model. MSE measures the mean of the squares of the errors, meaning it calculates the average squared difference between predicted and target values. Finally, MAE measures how far predicted values are from observed values with their absolute difference. Equations shown below:
\begin{equation}
    R^2 = 1- \frac{\sum_i(y_{i} - \hat{y_i})}{\sum_i(y_{i} - \bar{y})},
\end{equation}

\begin{equation}
    MSE = \sum_{i=1}^n(y_{i} - \hat{y_i})^2,
\end{equation}

\begin{equation}
    MAE = \frac{\sum_{i=1}^n|y_{i} - \hat{y_i}|}{n},
\end{equation}
where $n$ is the number of dataset, $\hat{y_i}$ is the predicted value, $y_i$ is the ground truth value, and $\bar{y}$ is the mean value. $\hat{y_i}$, in our case, is the similarity score of two inputs of segments, and $y_i$ is our ground truth of similarity score based on \textit{range of motion}, \textit{stability}, or \textit{repetition}.
 To test our model with the baselines, we evaluate the experiments on our local machine with a CPU of AMD Ryzen 9 5950X with a 16-Core processor (3.40 GHz), RAMs of 64 GB (3200 MHz), and a GPU of Nvidia GeForce RTX 3090. The operating system is Windows 10 Pro. Additionally, we envision to deploy our deep learning model into cloud server to support API call directly from our mobile application.

\subsection{Performance}\label{evaluation:subsec:performance}

\subsubsection{Siamese Similarity}
We explore different ways of quantitatively measuring our exercise metrics. There are two necessary baseline models: RNN and CNN because RNN is known for arbitrary input with time-varied data, and CNN learns well in spatial and temporal features \cite{stromback2020mm}. Additionally, we choose SimCLR as third baseline because its architecture enables useful representation learning in contrastive learning \cite{SIMCLR2020}. Thus, we provide and twist the networks such as SimCLR \cite{SIMCLR2020}, VGG, and RNN as our baselines to further understand our problem. In SimCLR, the original paper proposed using the hidden features in the final layer of ResNet as a representation to compare with a different image (or augmentation of the same images) to learn the underlying knowledge of images with a contrastive loss. However, in our problem, we return a label with a value between 0 and 1. We simplify the output procedure in SimCLR by having a Cosine Similarity as the final layer to compare the two images (in our case, two segments of signals) and add a Sigmoid activation function to return such a result. Additionally, since SimCLR utilizes ResNet in image comparison, we use their structure with a 1-D convolutional layer instead of 2-D to perform an equivariant operation as well as to max pooling procedure. Moreover, VGG is applied with similar 1-D convolution and 1-D max pooling to work with signals. Lastly, we also used a vanilla RNN as our baseline. We use the outputs of the last sliding window as hidden representations and feed into a similar output procedure as described earlier in SimCLR. 

\begin{figure}
    \centering
    \includegraphics[width=\textwidth]{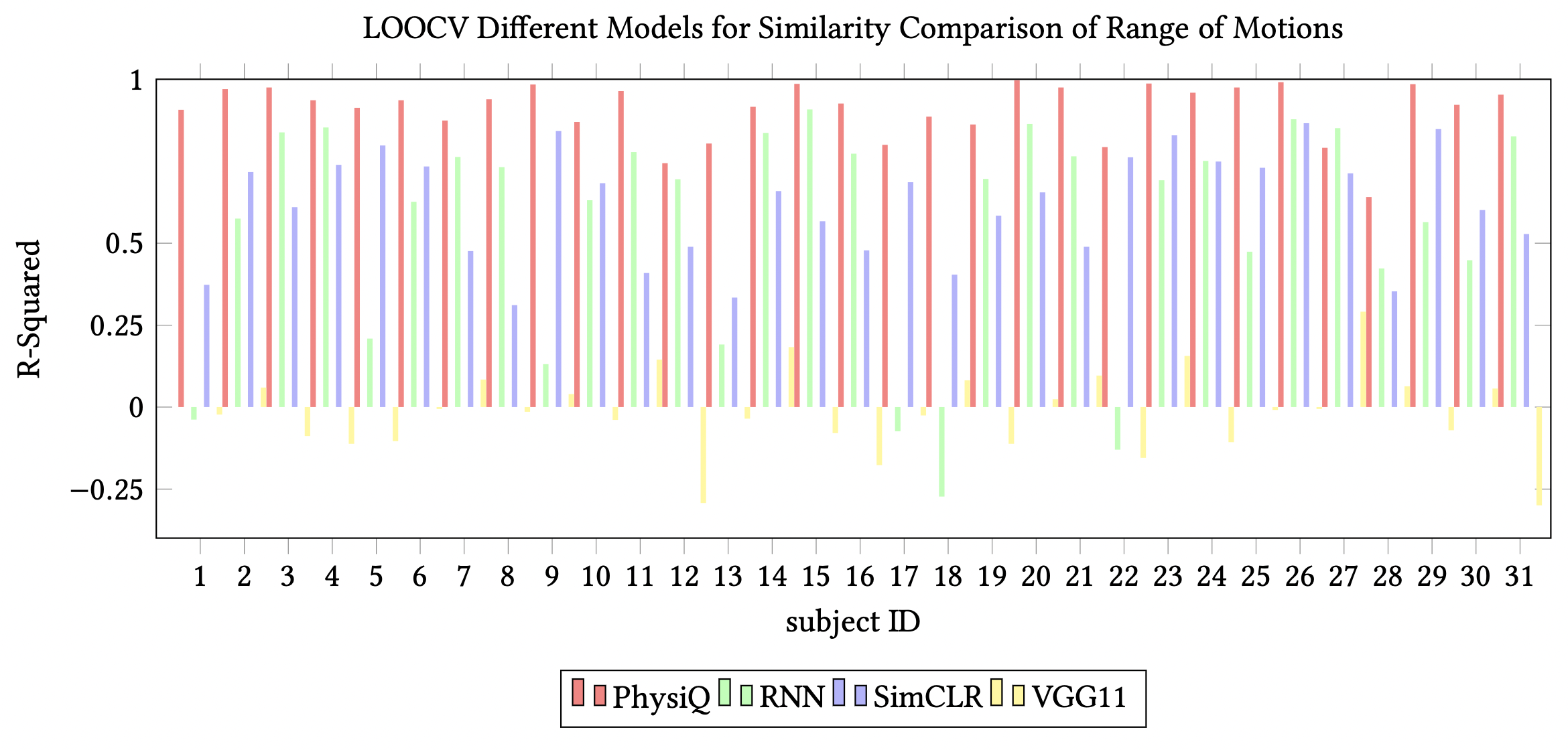}
\caption{This figure demonstrates that our framework, PhysiQ, outperforms many other architectures by capturing the spatiotemporal information. As shown above, the red line represents our framework, PhysiQ with R-Squared result on y-axis. Noted, the higher the R-Squared, the better the result as it explains how fitted our model is given the unseen data. Moreover, interestingly, as shown in green line, vanilla RNN captures the temporal information very well but it has problems of generalizability as in subject ID 1 while some of subjects the RNN model can predict very well.}
\label{figure:loocv_similar}
\end{figure}


There is a potential drawback between half arm span and half-half arm span exercises. Because of the limited external rotation motion, the model might have a harder time distinguishing all three metrics. But overall, our model in all three exercises outperforms the baseline extensively in the metrics of \textit{range of motion}, \textit{stability}, and \textit{repetition}, as shown in Table \ref{table:sa_701020}, Table \ref{table:er_701020}, Table \ref{table:ff_701020}. Moreover, because external rotation does not have as many waypoints of motion as forward flexion (external rotation only has three, but forward flexion has five), its results are not as good as shoulder abduction and forward flexion in \textit{stability}. Additionally, RNN and SimCLR can have a related good performance throughout the metrics but have difficulties with higher accuracy because of the low complexity of the models.
Similarly, in Figure   \ref{figure:loocv_similar}, the baselines have inconsistent performance throughout subjects, but PhysiQ outperforms them and consistently results well.

\begin{table}[h!]

\centering
\scriptsize

\scalebox{1.05}{
\begin{tabular}{||c || c |c|c|c||c|c|c|c||c|c|c|c ||}
\hline
\multicolumn{1}{||c||}{} & \multicolumn{4}{|c||}{ROM} & \multicolumn{4}{|c||}{Stability} & \multicolumn{4}{|c||}{Repetition} \\
 \hline
 
  &PhysiQ & SimCLR &RNN&VGG & PhysiQ & SimCLR &RNN&VGG  & PhysiQ & SimCLR &RNN&VGG  \\ [0.5ex] 
 \hline\hline
 MSE  & \textbf{0.00217  } &0.0153& 0.0138     &0.0442      &\textbf{0.00454} &0.0143&0.0149 &0.0561       &\textbf{0.00716}  &0.0145&0.0180&0.106\\ 
 MAE  & \textbf{0.0310  } &0.0927& 0.0914     &0.175        &\textbf{0.0514} &0.0930& 0.0966 &0.184        &\textbf{0.0690}   & 0.0941&0.102&0.272\\ 
 R-Square & \textbf{0.949  } &0.634& 0.676    &-0.0341       &\textbf{0.791} &0.342 & 0.314 &-0.0645         &\textbf{0.869}   &0.736&0.668&-0.931\\ 
 \hline
\end{tabular}}
\caption{The overall evaluation on exercise \textbf{shoulder abduction} similarity comparison of \textit{range of motion}, \textit{stability}, and \textit{repetition}.}
\label{table:sa_701020}
\end{table}

\begin{table}[h!]
\centering
\scriptsize

\scalebox{1.05}{
\begin{tabular}{||c || c |c|c|c||c|c|c|c||c|c|c|c ||} 
\hline
\multicolumn{1}{||c||}{} & \multicolumn{4}{|c||}{ROM} & \multicolumn{4}{|c||}{Stability} & \multicolumn{4}{|c||}{Repetition} \\
 \hline
 
  &PhysiQ & SimCLR &RNN&VGG & PhysiQ & SimCLR &RNN&VGG  & PhysiQ & SimCLR &RNN&VGG  \\ [0.5ex] 
 \hline\hline
 MSE  & \textbf{0.0117  } &0.0174& 0.0129& 0.0192       &\textbf{0.00166} &0.0121&0.00220&0.00291   &\textbf{0.0105}  & 0.0277 &0.0305&0.119\\ 
 MAE  & \textbf{0.0882  } &0.0990& 0.0937& 0.122        &\textbf{0.0250}  &0.0537&0.0309 &0.0339      &\textbf{0.0749}  & 0.134 &0.140&0.285\\ 
 R-Square & \textbf{ 0.757  } &-0.0404& 0.226&0.0184    &\textbf{0.423}   &-3.412&0.217  &-0.0236       &\textbf{0.805}   & 0.488 &0.437&-1.298\\ 
 \hline
\end{tabular}}
\caption{The overall evaluation on exercise \textbf{external rotation} similarity comparison of \textit{range of motion}, \textit{stability}, and \textit{repetition}.}
\label{table:er_701020}
\end{table}

\begin{table}[h!]
\centering
\scriptsize

\scalebox{1.05}{
\begin{tabular}{||c || c |c|c|c||c|c|c|c||c|c|c|c ||} 
\hline
\multicolumn{1}{||c||}{} & \multicolumn{4}{|c||}{ROM} & \multicolumn{4}{|c||}{Stability} & \multicolumn{4}{|c||}{Repetition} \\
 \hline
 
  &PhysiQ & SimCLR &RNN&VGG & PhysiQ & SimCLR &RNN&VGG  & PhysiQ & SimCLR &RNN&VGG  \\ [0.5ex] 
 \hline\hline
 MSE  & \textbf{0.00215  } &0.00963& 0.00988&0.0454           &\textbf{0.00602}    &0.0176 &0.0155   &0.0545             &\textbf{0.00276}   &0.0121  &0.0121  &0.0597\\ 
 MAE  & \textbf{0.0366  } &0.0771& 0.0673& 0.179            &\textbf{0.0594}     &0.103 &0.0948   &0.190              &\textbf{0.0411}   & 0.0888 &0.0888&0.200\\ 
 R-Square & \textbf{ 0.950  } &0.775& 0.772&-0.0515        &\textbf{0.883}      &0.657  &0.700    &-0.0450            &\textbf{0.9513}   &0.785   &0.785 &-0.0472\\ 
 \hline
\end{tabular}}
\caption{The overall evaluation on exercise \textbf{forward flexion} similarity comparison of \textit{range of motion}, \textit{stability}, and \textit{repetition}}
\label{table:ff_701020}
\end{table}

\subsubsection{Classification}
Next, we evaluate the PhysiQ's classification component on the quality of activity, i.e., how accurate can our model predict on \textit{ROMs}. Thus, our baselines for the models are some of the best classification models that modern machine learning and deep learning can achieve: CNN with Multi-Layer Perceptrons (MLP), LSTM with MLP, and Logistic Regression (Linear). Because classification and similarity comparison are two different problems, we target our problem using a different baseline but similar structure overall. CNN with MLP utilizes the 1-D Convolutional Neural Network to capture spatial knowledge and use the hidden representation to go through an MLP network to classify \textit{ROMs} and \textit{stability}. Similarly, LSTM with MLP has a similar approach, except a sliding windows segmentation is used to create a temporal representation, which feeds into the fully connected layer as MLP. Lastly, Logistic Regression is simply a fully connected model that considers all the points (dimensional flattening) in the signal and predicts the labels.

\begin{figure}
    \centering
    \includegraphics[width=\textwidth]{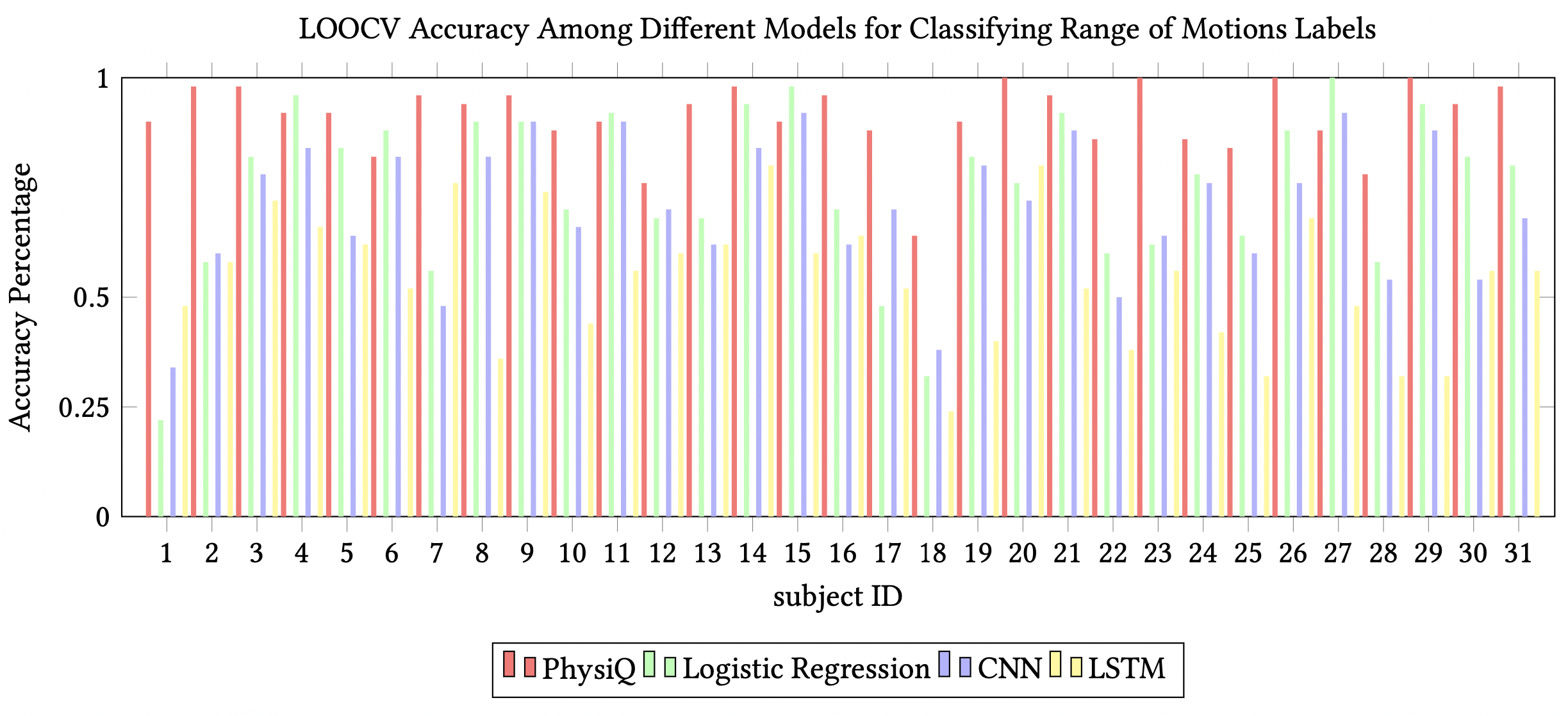}
\caption{This figure shows the accuracy of our model with output of classification of \textit{range of motion}. As showed in this figure, our PhysiQ (shown in red) performs mostly better than the other models in this diagrams with a percentage accuracy on y-axis.}
\label{figure:loocv_class}
\end{figure}

\begin{figure}
 \begin{subfigure}{0.24\textwidth}
     \includegraphics[width=\textwidth]{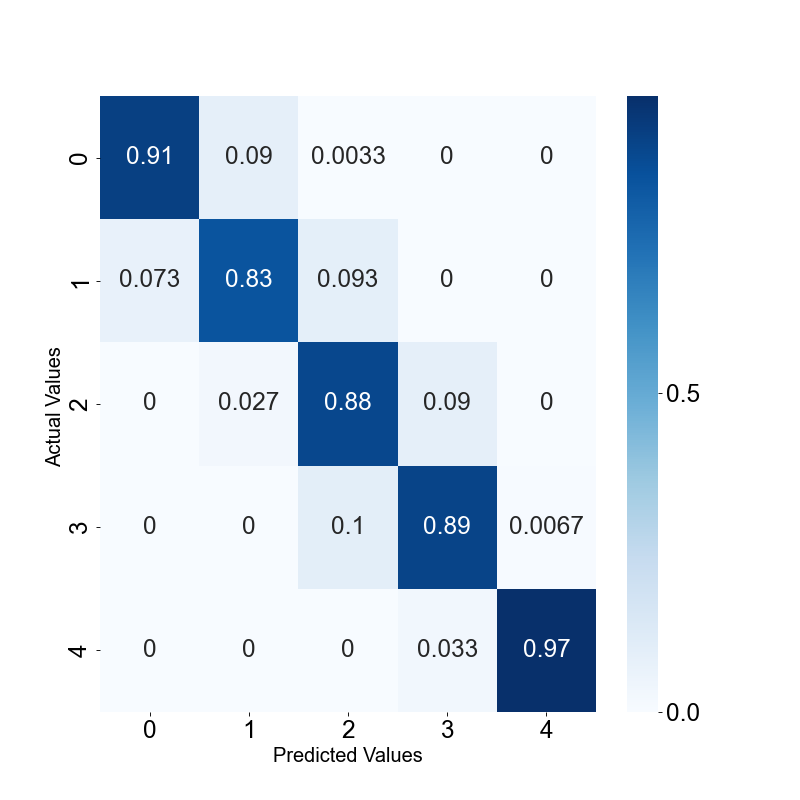}
     \caption{PhysiQ}
     \label{fig:sa_rom_classification:a}
 \end{subfigure}
 \hfill
 \begin{subfigure}{0.24\textwidth}
     \includegraphics[width=\textwidth]{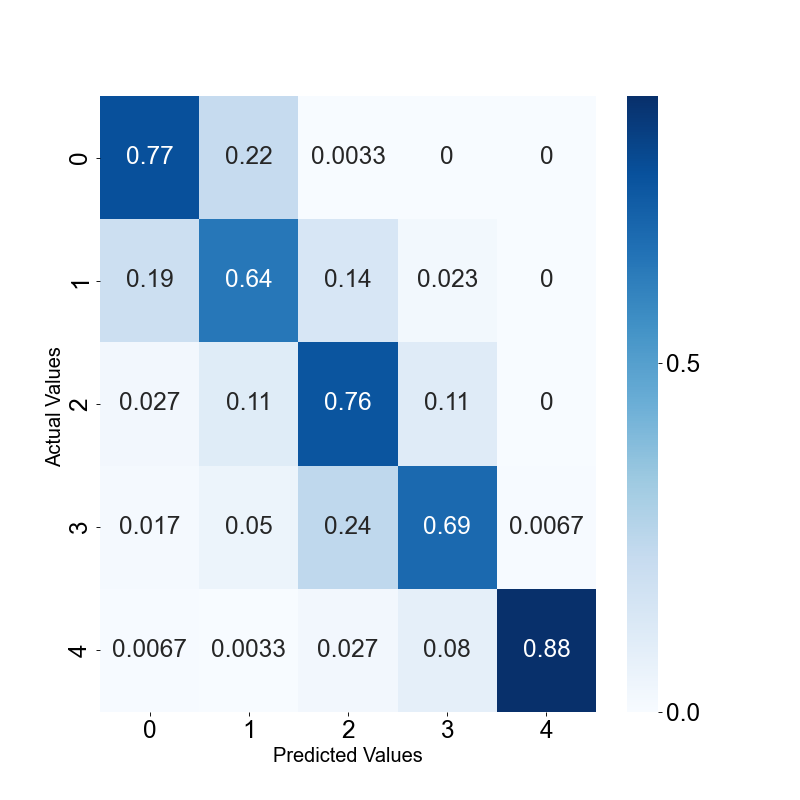}
     \caption{Logistic regression}
     \label{fig:sa_rom_classification:b}
 \end{subfigure}
 \hfill
 \begin{subfigure}{0.24\textwidth}
     \includegraphics[width=\textwidth]{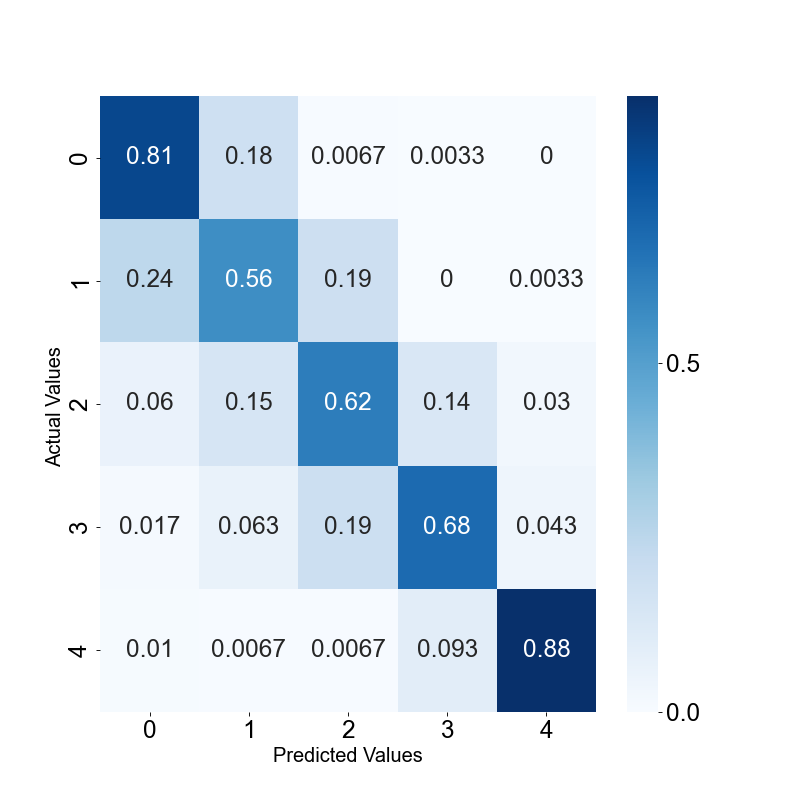}
     \caption{CNN}
     \label{fig:sa_rom_classification:c}
 \end{subfigure}
 \hfill
 \begin{subfigure}{0.24\textwidth}
     \includegraphics[width=\textwidth]{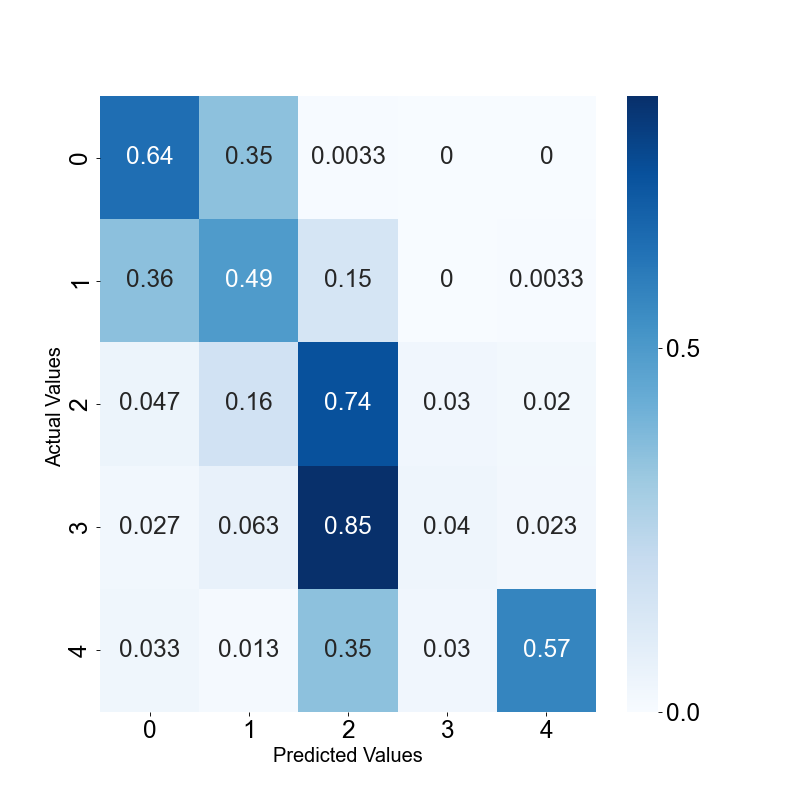}
     \caption{RNN}
     \label{fig:sa_rom_classification:d}
 \end{subfigure}

 \caption{This is the confusion matrices for classification of \textit{range of motion} in \textbf{shoulder abduction} with PhysiQ and baselines. Shoulder abduction has class of 5 from 30, 60, 90, 120, and 150, which is labeled from 0 to 4}
 \label{fig:sa_rom_classification}

\end{figure}

\begin{figure}
 \begin{subfigure}{0.24\textwidth}
     \includegraphics[width=\textwidth]{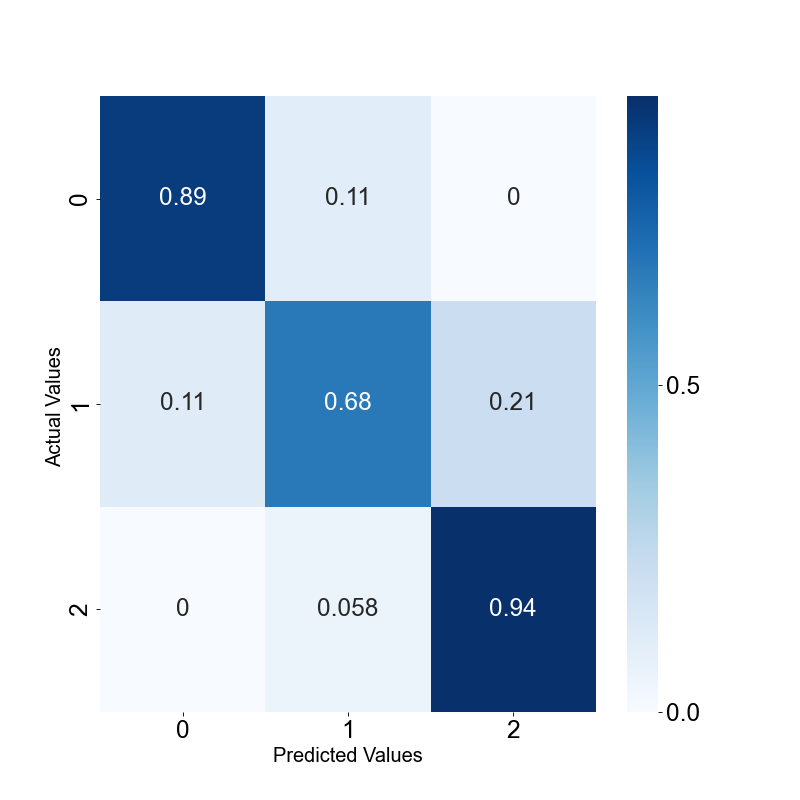}
     \caption{PhysiQ}
     \label{fig:er_rom_classification:a}
 \end{subfigure}
 \hfill
 \begin{subfigure}{0.24\textwidth}
     \includegraphics[width=\textwidth]{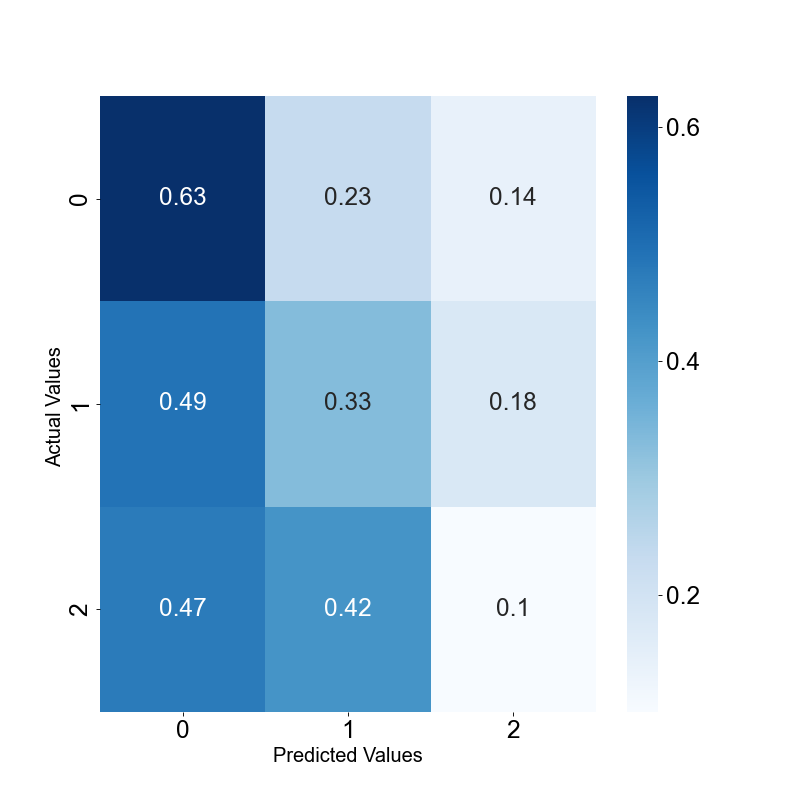}
     \caption{Logistic regression}
     \label{fig:er_rom_classification:b}
 \end{subfigure}
 \hfill
 \begin{subfigure}{0.24\textwidth}
     \includegraphics[width=\textwidth]{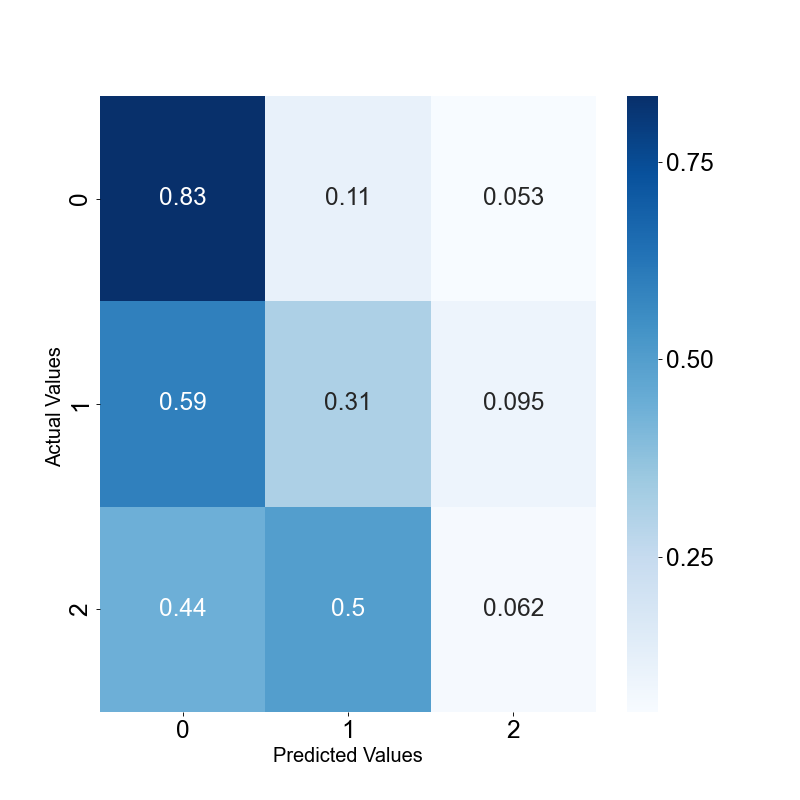}
     \caption{CNN}
     \label{fig:er_rom_classification:c}
 \end{subfigure}
 \hfill
 \begin{subfigure}{0.24\textwidth}
     \includegraphics[width=\textwidth]{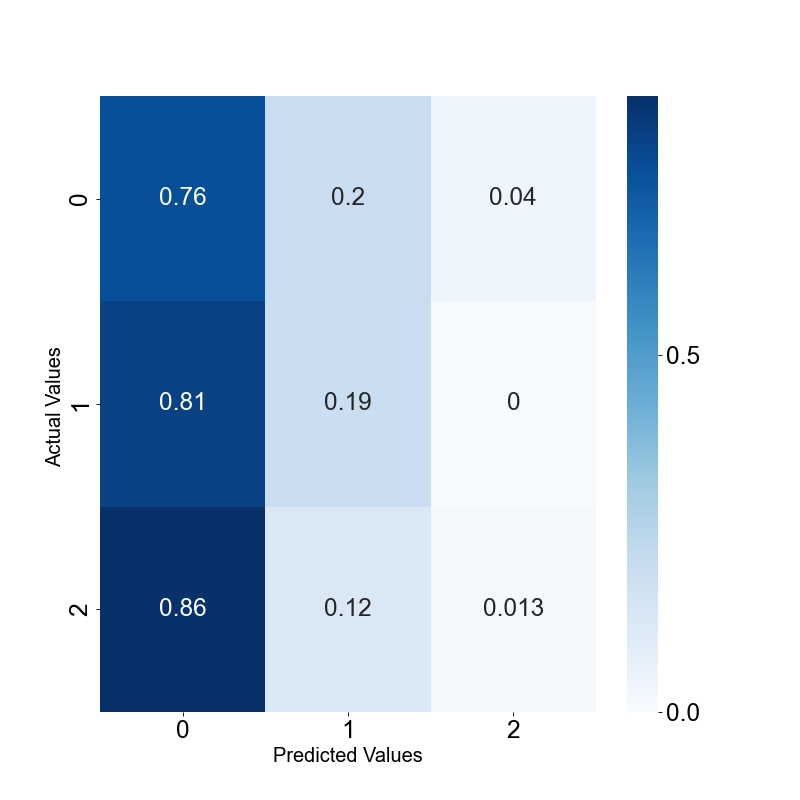}
     \caption{RNN}
     \label{fig:er_rom_classification:d}
 \end{subfigure}

 \caption{This is the confusion matrices for classification of \textit{range of motion} in \textbf{external rotation} with PhysiQ and baselines. The external rotation has a class of 3 from 45, 90, and 150, which is labeled from 0 to 2}
 \label{fig:er_rom_classification}

\end{figure}

\begin{figure}
 \begin{subfigure}{0.24\textwidth}
     \includegraphics[width=\textwidth]{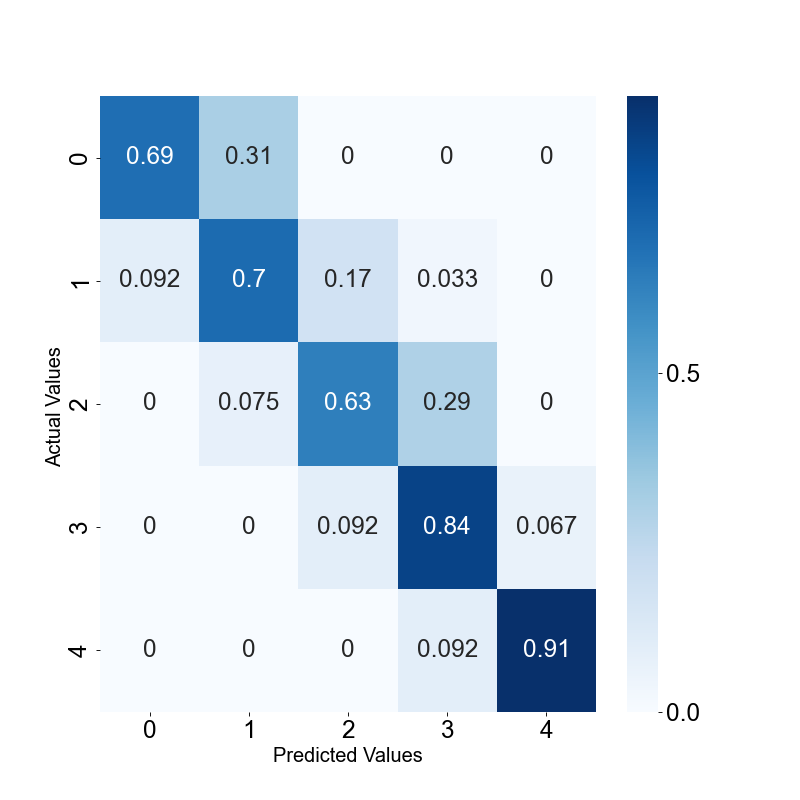}
     \caption{PhysiQ}
     \label{fig:ff_rom_classification:a}
 \end{subfigure}
 \hfill
 \begin{subfigure}{0.24\textwidth}
     \includegraphics[width=\textwidth]{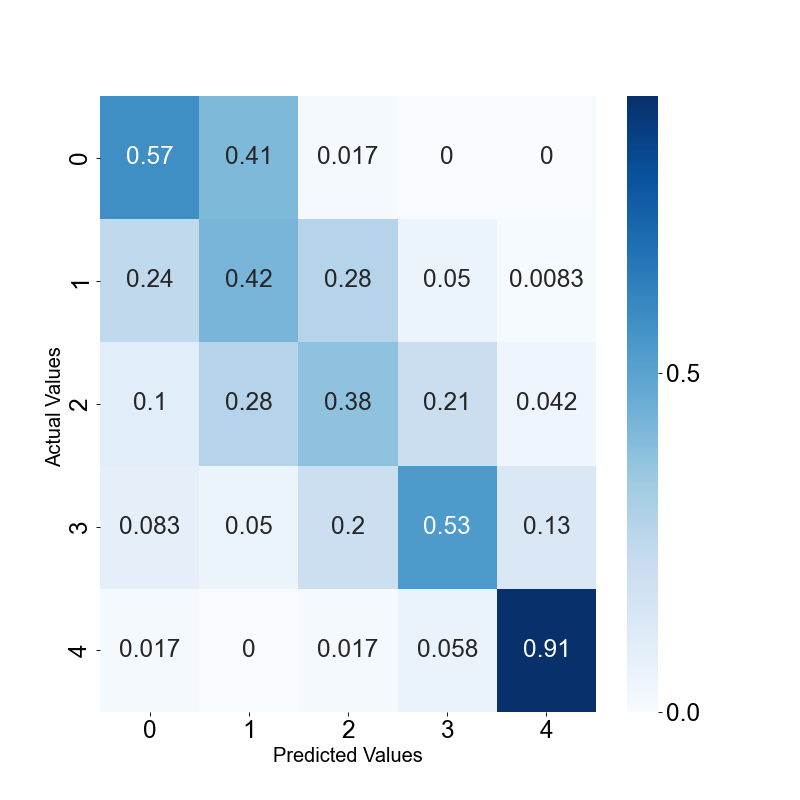}
     \caption{Logistic regression}
     \label{fig:ff_rom_classification:b}
 \end{subfigure}
 \hfill
 \begin{subfigure}{0.24\textwidth}
     \includegraphics[width=\textwidth]{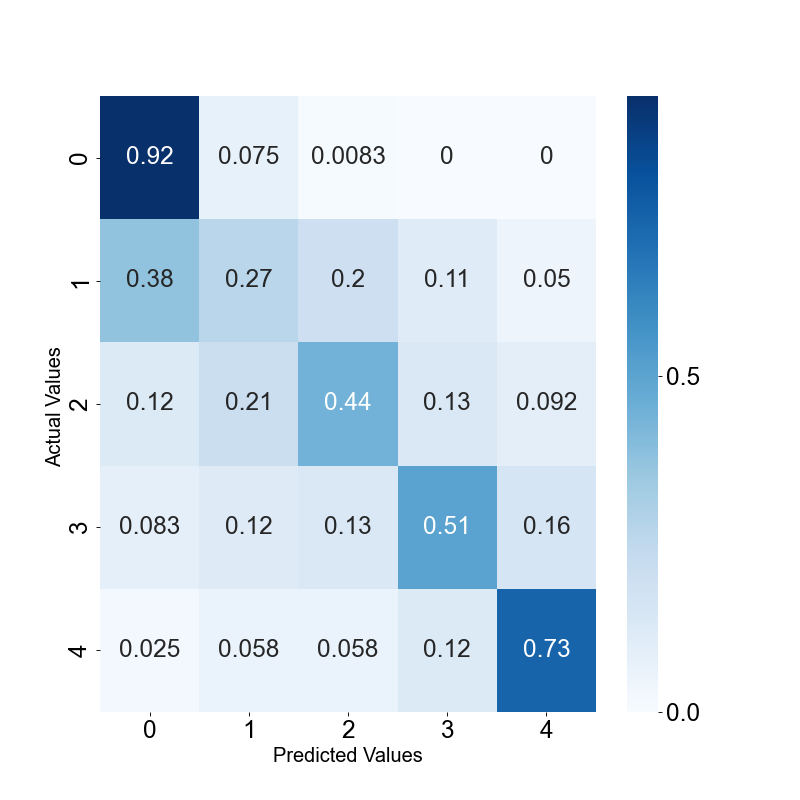}
     \caption{CNN}
     \label{fig:ff_rom_classification:c}
 \end{subfigure}
 \hfill
 \begin{subfigure}{0.24\textwidth}
     \includegraphics[width=\textwidth]{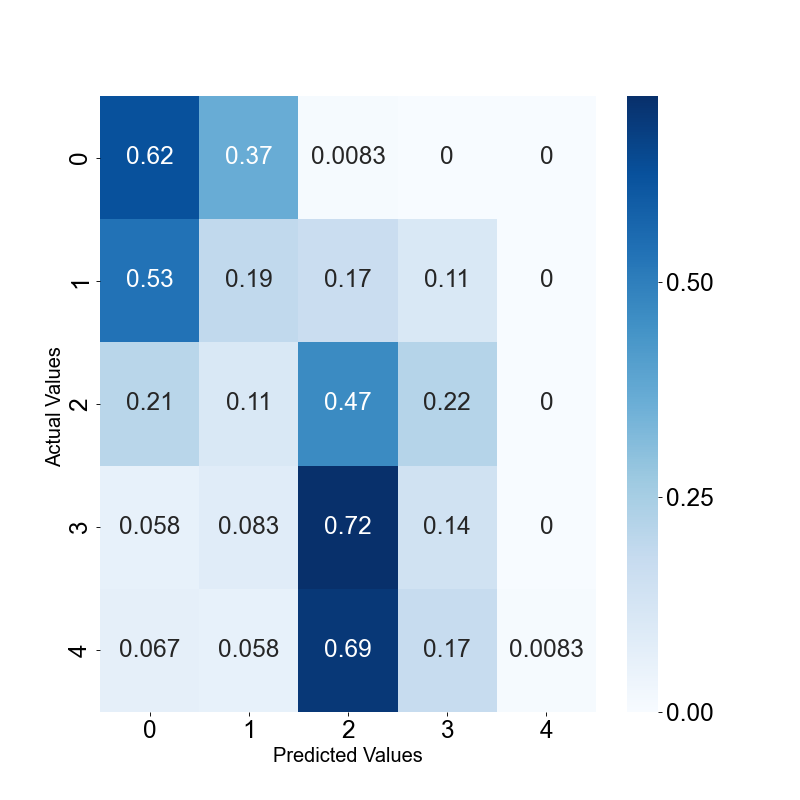}
     \caption{RNN}
     \label{fig:ff_rom_classification:d}
 \end{subfigure}

 \caption{This is the confusion matrices for classification of \textit{range of motion} in \textbf{forward flexion} with PhysiQ and baselines. Forward flexion has a class of 5 from 30, 60, 90, 120, and 150, which is labeled from 0 to 4}
 \label{fig:ff_rom_classification}

\end{figure}

As a result, we provide confusion matrices for all baselines and our model for classification of \textit{range of motion}. As shown in Figure   \ref{fig:sa_rom_classification:a}, \ref{fig:er_rom_classification:a}, and \ref{fig:ff_rom_classification:a}, compared against the other 3 baselines, our model demonstrates the best performance overall with a minimal inaccuracy on higher degree \textit{range of motion} exercises. The difficulties in differentiating between 30 and 60 degrees happen because of limited supervision and little self-reflection (such as a mirror), and all subjects might not perform homogeneously in the exercises. As a result, this might leads to inconsistency in the model to understand the difference between 30 degrees and 60 degrees ROM.


%

\subsection{Parameters Evaluation}\label{evaluation:subsec:parameters}
\subsubsection{Sliding Windows}
To evaluate the accuracy of our proper performance, we use different sliding window segmentation of samples to test our model results. We test the performance on leave-one-subject-out cross-validation (LOOCV). By having sliding windows that are 50, 100, and 150. The model decreased its performance when the sliding windows increased while keeping the same step size. Increasing the sliding window size decrease the number of time, or redundancy, for the model to see. Having smaller sliding windows helps the model to understand the quality of exercises. At the same time, as we increase the step size (the size of overlapping), the model also performs better but sees a drop in the testing dataset. An extensive redundancy overfits the training distribution and does not generalize it well. As a result, we choose a sliding window of 50 with a step size of 15 for training, validating, and testing, which alike to our data collection rate of 50 Hz. Having some redundancy helps the model to generalize the relationship between sliding windows.

\subsubsection{Repetition Size}
In repetition size, we have a different size in similarity comparison. We have each ID number of segmented exercises for each subject. Adjacent ID number represents that they can be concatenated together to output different. We use the same number of repetition sizes in training and testing. The model has a slight difference in accuracy (3 - 5 percent) in bigger repetitions as length increases. However, this is also affected by the hyperparameters that we choose. Without changing the step size and window size, the model can still perform well because our model generalizes very well. Moreover, we also see a performance increase when adding dropout regularization deep LSTM network. As a result, the performance of four repetitions becomes 0.921 in the R-Squared correlation.

\subsubsection{Padding}
We also evaluate the effect of front padding and back padding. We hypothesize that there should be no difference between front and back padding. However, when the padding is at the end of the temporal sequence, the model does not learn, resulting in an average R-Squared score of 0. Moreover, the results remain unchanged throughout the epochs. This non-learning behavior happens possibly due to the fact that many of the segments are varied significantly. In that cases, many of the windows remain zero and do not help the model to learn. At the same time, as we attempt the model without padding or minimize the amount of padding, we are facing the issue that the temporal model could cheat the result because there is a correlation between higher \textit{range of motion} exercises and the length of the data. In contrast, lower ROM exercises can have a shorter overall length. This relation could result in an accumulating effect of shortcuts for the deep learning model to learn and cheat. So using constant front padding for all one repetition exercises is consistent and performs similarly in a different number of repetitions.  
\subsection{Ablation Study}\label{evaluation:subsec:ablation}
\subsubsection{Removal of Hidden Representations} 
After removing the \textbf{temporal representation}, LSTM encoder, part of PhysiQ, the model still performs relatively well with an MSE 0.007168, MAE 0.0665, and R-Squared 0.8346 with 256 hidden features. At the same time, we did the same procedure for \textbf{spatial representation}. After removing the CNN encoder from PhysiQ, the model performs relatively well with an MSE 0.00873, MAE 0.063, and 0.8208 in R-Squared with one hidden temporal layer and no dropout regularization. Interestingly, in the usages of both spatial and temporal models, our model can achieve 0.92 R-Squared while removing either of them drops significantly by around 10 percent. The decrease happened due to the reasoning that the signal data contain both spatial and temporal information, and a single model can only capture part of the information with the help of an attention mechanism. 
\subsubsection{Dropout}
Testing the dropout is also essential to see how the model is generalized. This dropout is applied in every layer, including the spatial encoder (CNN) and temporal encoder (LSTM). We used LOOCV to compare our results with 0\%, 20\%, and 50\% dropout rates to see if the difference is significant. We test our model with hidden features of 256 for temporal and spatial encoding, two LSTM layers, 16 heads of attention, and a batch size of 1024. The model should not have a significant difference between 20\% and 50\% but possibly a slight performance boost in 20\% or 50\% compared to 0\%. Our model is meant to generalize well across participants and should not overfit the training distribution; the average result of 0 percent dropout across the participants is 89.66\%. The average result with 20 percent dropout is 89.73\%. The average result with 50 percent dropout is 87.87\%. Interestingly, the 20\% dropout rate generates a slightly better results than 50\%.  This is because 20\% has already created its maximal regularization for the model, and increasing the dropout rate does not benefit from regularization. On the other hand, a much higher dropout rate could result in a higher variance in the model, resulted degradation in performance. Having only a 20\% dropout rate creating the maximum of regularization makes our model more convincing in terms of generalizability. 

\subsubsection{Attention}
We test the importance of attention in our model. Attention is applied to selectively focus on more critical aspects of the input sequence. In our model, attention is a final layer to measure the relationship between each hidden state. With that in mind, we remove the attention layer to compare results in the metrics of the \textit{range of motion} in shoulder abduction, external rotation, and forward flexion. Without the attention layer, the performance of LOOCV in shoulder abduction drops to a 0.892 R-squared correlation from 0.908. On exercise external rotation, the performance of LOOCV drops to a 0.560 R-squared correlation from 0.675. Meanwhile, the performance of LOOCV on forward flexion has an insignificant increase from 0.815 to 0.830. In summary, if the motion data is distinguishable and representative, the attention layer does not have a huge impact. However, when exercise has a limited motion, such as half-half arm span exercise external rotation, the attention layer is more impactful.

\section{A Survey on User Experience}
\label{sec:survey}

To investigate PhysiQ's application in practice, we surveyed the participants who used our system for data collection. Users wear an Apple Watch with a connected iPhone with our PhysiQ apps installed during data collection. In the beginning, users have the exercise instruction on the phone to start. 
PhysiQ on the phone visualizes the signal to see the repetition quality and feedback to the users (as shown in Fig. \ref{fig:PhysiQ_GUI}). 

After collecting the data, we distribute a questionnaire to all users. The questionnaire asks six questions on different scales. 
We designed the survey questions for three purposes. First, we get users' feedback on our current design (Q1). Secondly, we explore users' preferences in alternative platforms and recommendations to improve our current design (Q2 and Q3). Last, we gather users' demographic information (presented in Section \ref{sec:data collection}) and ask behavioral questions (Q4-Q6). We study the correlation between users' behaviors and the performance of our algorithms in measuring their activities. 
We attach the questions and summary of results below because we believe our findings are valuable for our future work and the research community. 

\begin{itemize}
    \item \textbf{Q1: How do you like the current feedback and recommendation system?} 
        \begin{itemize}
        \item scale of 1 to 5, with 1 being the lowest, and 5 being the highest
        \end{itemize}
    \item \textbf{Q2: If we have a different platform, what platform will you like to have recommendation system of the exercises on?}
        \begin{itemize}
        \item Smartwatch
        \item Smartphone
        \item Smartglasses such as VR, AR glass
        \end{itemize}
    \item \textbf{Q3: During what period of the exercise do you like such feedback? For example, for today's exercise session, you have 5 different exercises to perform and for each exercise, you have 10 repetitions of 5 sets? }
        \begin{itemize}
        \item After a set of exercises  (during this particular exercise, after 10 repetitions)
        \item After the particular exercise of 5 sets
        \item After the entire exercise session (after 5 exercises)
        \end{itemize}
    
     \item \textbf{Q4: During the time of collecting your exercise data, do you drink coffee or any caffeinated drinks regularly? If so, how often? }
        \begin{itemize}
        \item Every day
        \item A few times a week
        \item About once a week
        \item A few times a month
        \item Once a month
        \item Less than once a month
        \end{itemize}
    
    \item \textbf{Q5: During the time of collecting your exercise data, how many hours do you sleep? } 
        \begin{itemize}
        \item Time in hour
        \end{itemize}
        
    \item \textbf{Q6: This question is regarding your medical history and you do not need to specify the medication. At the time of collecting your exercise data, do you know and take any medication at the time that would affect your ability to perform exercise or activity? }
        \begin{itemize}
        \item Yes or no
        \end{itemize}
    \end{itemize}
    
Out of 31 participants, we have 27 participants who complete the follow-up survey. Overall, we have an average of 4.26 rating on how much the users like our system. Interestingly, we have 48.1\% of the users who want to have a recommendation on a smartwatch, 51.9\% of the users on a smartphone, and no one wants to use smart glass to get recommendation feedback. Additionally, 59.3\% of our users want to have their feedback after a set of exercises, 22.2\% after the particular exercises, and 14.8\% after an entire session. Lastly, one participant suggests they should be able to see the feedback anytime they want.

Moreover, we have 18.5\%, 33.3\%, 14.8\%, 18.5\%, 3.7\%, and 11.1\%, respectively, on the frequency of coffee consumption based on the answer order above. At the same time, on average, our participants sleep 7.63 hours with a minimum of 6 and a maximum of 9. Lastly, we only have 1 participant possibly on medication that affected their performance overall, but we do not find that medication was affecting the performance of our model testing on this participants' exercises.

\section{Discussion}
\label{sec:discussion}

\subsection{Applications}
\emph{Physical Therapy Home Assessment}. This application is intended to work for patients and people injured, postoperative, or mentally traumatized. While performing at-home exercises, our application can provide real-time feedback and assess the quality of the exercises to provide better interaction and supervision while clinics are not accessible. At the same time, we believe our model is capable of analyzing and predicting people who might suffer from a different illness or potential injuries. One example is that people with heavy usage of handcrafting might suffer from carpal tunnel syndrome, which is caused by pressure on the median nerve. 

\emph{Daily Exercises Assessment}. This application can also support working with people who enjoy exercise. People can benefit from it by closely assessing how they perform specific repetitions of exercises. Moreover, this application could also apply to people who are playing sports. We envision that our model can eventually support people who play sports like tennis to predict a player's direction, speed, or posture.

\subsection{Assumption and Limitation}
This paper discusses the potential of using a smartwatch to support users assess their exercises with a quantitative measure of the quality. However, there are a few assumptions made to support this application. First of all, we assume that the users' posture should have some level of decency. For example, suppose users perform the exercise of shoulder abduction while bending their neck or wiggling around as not in a straight form. In that case, the application likely can still give a good score on the exercises simply because the users might still perform the exercise correctly. Because the smartwatch cannot capture all the movement within the body, we do not have the luxury of analyzing all the posture. Secondly, 
we assume that the primary segmentation is performed upon accelerometer data, and its energy is only extracted from the accelerometer. Thirdly, 
we assume there is 5 level of quality of exercises in \textit{range of motion} in shoulder abduction. These are our metrics to digitalize from body movement to computational numbers. The five levels can vary based on different professionals' metrics, and our goal is to standardize a metric in our model that can measure and understand.

We notice that the energy plot does not have a clear pattern in some exercises for some subjects. Such limitation is due to our little knowledge regarding the data collection. Applying our energy plot when collecting data from the users could be more rigorous and use it as a checkpoint to verify if the subjects have correctly performed the exercises.
In the future work, we will improve our framework to better handle above assumptions and limitations.  

\section{Summary}
\label{sec:summary}


We develop an innovative system, PhysiQ, to quantitatively digitalize the quality of exercises through new metrics on a commodity smartwatch. We scrutinize and verify that different metrics and exercises have unique characteristics that can be recognized and understood by our deep learning model. By developing such a model based on Siamese Neural Network with additional spatiotemporal representation encoding, our model can achieve 95 percent R-square correlation and 90 percent accuracy in classification. Moreover, a comprehensive evaluation and user studies are performed to show the effects on our metrics in range of motion, stability, and repetition. 
The end goal is to improve the prediction and assessment of people who needs therapy to improve their quality of life. In addition, we envision that current technologies and relevant professions can be benefited from it by expanding usability, generalizability, and model learnability. By combining deep learning and the physical therapy method, we believe that our framework is the tool to lead people to improve their quality of life.

\section{Acknowledge}
This work was supported in part by National Science Foundation 2220401. We would also like to thank Zirong Chen, Xing Yao, David Atwood, Melissa Wang, Anna Chen, and Yashvitha Thatigotla for their effort to help and discuss this project.

\bibliographystyle{ACM-Reference-Format}
\bibliography{sample-base}

\end{document}